\documentclass[%
reprint,
superscriptaddress,
groupedaddress,
runinaddress,
showpacs,preprintnumbers,
bibnotes,
amsmath,amssymb,
aps,
prb,
floatfix,
]{revtex4-1}

\usepackage{graphicx}
\usepackage{dcolumn}
\usepackage{dsfont}
\usepackage{bm}
\usepackage{bbm}
\usepackage{hyperref}
\usepackage[mathlines]{lineno}

\usepackage{graphicx}

\begin{document}
\title{Radiative Heat Transfer in Fractal Structures} 

\author{M. Nikbakht}
\email{mnik@znu.ac.ir}
\affiliation{Department of Physics, University of Zanjan, Zanjan 45371-38791,Iran.}

\date{\today}
\begin{abstract}
The radiative properties of most structures are intimately connected to the way in which their constituents are ordered on the nano-scale. We have proposed a new representation for radiative heat transfer formalism in many-body systems. In this representation, we explain why collective effects depend on the morphology of structures, and how the arrangement of nanoparticles and their material affects the thermal properties in many-body systems. We investigated the radiative heat transfer problem in fractal (i.e., scale invariant) structures. In order to show the effect of the structure morphology on the collective properties, the radiative heat transfer and radiative cooling are studied and the results are compared for fractal and non-fractal structures. It is shown that  fractal arranged nanoparticles display complex radiative behavior related to their scaling properties. we showed that, in contrast to non-fractal structures, heat flux in fractals is not of large-range character. By using the fractal dimension as a means to describe the structure morphology, we present a universal scaling behavior that quantitatively links the structure radiative cooling to the structure gyration radius.
\end{abstract}

\maketitle

\section{Introduction}\label{sec1}

The field of radiative heat transfer is of considerable interest due to its promise for non-contact modulation of heat transfer. In the past years, several efforts have been made to understand and analyze the radiative heat transfer at the nano-scales. It is well-known that the radiative transfer between two objects depends drastically on their separation distance, and at small separation distances compared to thermal wavelength, the flux is several orders of magnitude larger than the value predicted by Stefan-Boltzmann
law\cite{theorytoexperiment}. A major advance in the field was made by Polder and Van-Hove in the use of Rytov’s theory of fluctuational electrodynamics for describing radiative heat transfer at the nano-scales\cite{VanHove}.
  During the past few years significant attention has been paid on the influences of size \cite{twobodysize,twobodysize2,sizethreebody}, shape and relative orientation \cite{spheroidslab,threebodyshape}, and materials\cite{twobodymetal,materialslab,manybodymagnetic}, on the radiative heat transfer and thermal evolution \cite{dynamicsNikbakht,threebodydynamicmessina}, in two or three-body systems. These studies show that the significant enhancement in the heat transfer at small separation distances is due to the feature of the near-fields at this scale. 

The rapid growth of physical analysis methods and nano-fabrication techniques \cite{synthesis2,fabrication}, providing researchers with the necessary tools for designing and predicting setups in order to manipulate the radiative heat flux at larger systems.  Bringing more than two objects at small distances changes the radiative properties due to many-body effects\cite{manybodynikbakht,many-bodyben,manybodyzhao,manybodyslab}.  This fact arises from the multiple scattering of the radiation field by the objects in a system which accompanied by new modes participating in heat transfer\cite{threebodyshape,manybodyheatdiffusion}. Accordingly, the arrangement of nanoparticles in many-body systems has an important rule on the systems radiative properties. When these objects are widely separated, they can be regarded to scatter the radiation field independently. At higher volume fractions, or in case where fractal/periodic arrangement takes place, closer packing of nanoparticles influences the scattering of individual particles which can not be regarded to scatter the heat flux independently any more\cite{fractalscattering,localizemodes}.  In spite of major efforts, most of the theoretical considerations in the collective effects are restricted to few-body systems.   There are very few studies on the heat transfer problem in larger systems with ordered or disordered structure, including  ballistic and diffusive heat transfer in chain of nanoparticles\cite{arrayplasmonic,arraynanofluid,manybodyheatdiffusion}, energy and momentum transfer \cite{momentumheattransfermanybodyplanar} and ballistic regime of heat transfer \cite{ballesticslab} in many-body planar systems, heat transfer between cluster of nanoparticles \cite{htbetweencluster}, and heat transfer in  many-body dipolar systems with magnetic field\cite{magnetoplasmonicchain}. 

The majority of structures existing in nature turn out to be fractal\cite{vicsek1992fractal}. In contrast to ordered or disordered structures, fractal structures do not possess transnational invariance, accordingly, they can not transmit running waves\cite{opticalproperties1}.   In contrast, the coupling between nanoparticles is of long range in structures with lattice translation symmetry e.g., structures whose constituents are arranged in a highly ordered microscopic structure in one, two or three dimension. For the description of morph structures in terms of a limited set of parameters, a major improvement has been the introduction of fractal concepts by Mandelbort \cite{mandelbrot}. The fractal dimension reflects the internal morphology of the structure and depends on the rule which is used in building the fractal structure. The Diffusion Limited Aggregation (DLA) modeled by Witten and Sander \cite{wittensander}, and Cluster-Cluster Aggregation (CCA) modeled by Meakin \cite{meakinfractal2}, are examples of the use of the concept of fractal in simulation. These models involve growth of the structure by allowing nanoparticles (and subclusters) to diffuse and stick to the growing structures. Presently, various deterministic fractal structures can be artificially created due to a rapid progress in nanotechnologies. 
It is generally known that fractality has a strong influence on the optical properties of structures\cite{opticalproperties1,opticalproperties,nikbakht3}(specially in the case of nobel metal fractals), and mainly accompanied with inhomogeneous localization and strong enhancement field at some parts of these structures\cite{localizationstockman,localizemodes,localizedsurfaceplasmon}. Similar to optical responses, the radiative characteristics of fractal structures it expected to be different from those of ordered or disordered structures.

In this paper we present a novel approach for the analysis of radiative heat transfer problem in ensemble of nanoparticles, which allows one to describe how new behavior in thermal properties emerges from many-body interactions.  The proposed formalism is based on the representation of the heat transfer and radiative cooling of structures in terms of radiative modes (thermal excitations). This representation is used to analyze the transmission coefficient between nanoparticles in fractal and non-fractal structures. The calculations for fractals are restricted to the fractal structures based on Vicsek model \cite{vicsekfractal}. Moreover, silver nanoparticles which support surface plasmons are used as a typical material. We also restrict our self to dipolar regime (the separation distances are large compared to the nanoparticles sizes) in calculating heat flux, in part because multipolar interactions do not alter the qualitative feature of the phenomena. It is shown that the thermal conductance can be large even for far apart particles in structures showing a transnational symmetry. In particular, we have demonstrated that in contrast to non-fractal structures, the collective modes tend to be localized in fractal structures. Based on the scaling/transnational symmetries, it is concluded qualitatively that there exists maximum scale lengths which thermal radiation could effectively flow in fractal/ordered structures. Owing to this confined radiative diffusion area, we showed that the radiative cooling of structures possess a universal scaling properties.

The structure of the paper is as follow. The formalism is developed in Sec.~\ref{sec2}, where transmission coefficients, cooling coefficients, and conductance   are derived in terms of the eigen values and eigen vectors of the interaction matrix. In Sec.~\ref{sec3} we briefly introduces the Viscek fractal and its scaling properties. The radiative heat transfer in fractal and non-fractal structures are discussed in Sec.~\ref{sec4}.  Based on the interaction matrix representation, we calculated the transmission coefficient and mutual-conductance between particles in Sec.~\ref{sec4-1}. The same technique is applied for calculating radiative cooling of  structures, and the influence of the structure size on the cooling rate is investigated in Sec.~\ref{sec4-2}. Finally, our work is summarized in Sec.~\ref{sec5}. 

\section{Many-body radiative heat transfer formalism: geometric approach}\label{sec2}

Let us describe the basic ingredients of theoretical formalism we used to describe the radiative heat flux in many-body systems. The system under consideration consist an ensemble of $N$ distinct nanoparticle located at points ${\bm r}_i=(x_i,y_i,z_i)$, $i=1,\cdots,N$ inside a thermal bath at temperature $T_{b}$. For the sake of simplicity, nanoparticles are assumed to be identical spheres with radius $R$. Nanoparticles temperatures are $T_i$ and they are assigned a fluctuating dipole ${\bf P}_i^f$ representing their thermal radiation. Each nanoparticle receives the direct energy that radiated by other dipoles as well as the radiation energy that scattered between particles in the system. In the case of identical temperatures, i.e. $T_1=T_2=\cdots=T_N=T$, the net power exchange between two arbitrary particles in an ensemble would vanish. Accordingly, in the absence of thermal bath (i.e., $T_b=0$), the temperature evolution of the system is decided by the total power lost by all nanoparticle in an ensemble.   Nanoparticles exchange energy through dipolar interaction and the local electric field for a nanoparticle, located at ${\bm{r}}_i$ in the system, is determined by
\begin{equation}
{\bf E}_i=\sum _{j=1}^N{\hat{\bf G}}_{ij}{\bf P}_j~,
\label{eq1}
\end{equation}
where, $\hat{\bf G}_{ij}$ is a free space dyadic Green's tensor, gives the dipolar intraction between particles {\it i} and {\it j}. 
\begin{eqnarray}
\label{eq2}
&&\hat{\bf G}_{ij}=\frac{k^3}{4\pi}\Big[f(kr_{ij})\mathbbm{1}+g(kr_{ij})\frac{{\bf r}_{ij}\otimes{\bf r}_{ij}}{r_{ij}^2}\Big]\\
&&\nonumber f(x)=[x^{-1}+ix^{-2}-x^{-3}]\exp(ix)\\
&&\nonumber g(x)=[-x^{-1}-3ix^{-2}+3x^{-3}]\exp(ix)
\end{eqnarray}
where $k=\omega/c$, and $r_{ij}=|{\bf r}_{i}-{\bf r}_{j}|$ is the distance between $i$-th and $j$-th nanoparticles located at points ${\bf r}_{i}$ and ${\bf r}_{j}$, respectively. The Green's function has the contribution of near-, intermediate- and far-zone terms, $\propto r^{-3}$, $r^{-2}$ and $r^{-1}$ respectively. 
The term ${\hat{\bf G}}_{ij}{\bf P}_j$ in Eq.~(\ref{eq1}) gives the dipolar radiation (scattered and radiated) by particle {\it j} with dipole moments ${\bf P}_j$ at the point ${\bf r}_{i}$. The complex $3\times 3$ matrics $\hat{\bf G}_{ij}$ are symmetric, i.e., $\hat{\bf G}_{ij,\alpha\beta}=\hat{\bf G}_{ij,\beta\alpha}$ where Greek indices stand for the components. Moreover, $\hat{\bf G}_{ij}=\hat{\bf G}_{ji}$, The dipole moments can be represented in terms of the fluctuating and induced parts:
\begin{equation}
{\bf P}_i={\bf P}_i^I+{\bf P}_i^f,
\label{eq3}
\end{equation}

where ${\bf P}_i^I$ is the induced dipole moment and related to the local field through the relation
\begin{equation}
{\bf P}_i^I=\alpha\sum _{j\neq i}^N{\hat{\bf G}}_{ij}{\bf P}_j~,
\label{eq4}
\end{equation}
where $\alpha$ is the dressed polarizability tensor \cite{threebodydynamicmessina,radiationcorrection}
\begin{subequations}
\begin{eqnarray}
\alpha&=&\frac{\alpha_0}{1-G_\circ\alpha_0}~,\\
\alpha_0&=&3v\frac{(\epsilon-\epsilon_h)}{(\epsilon+2\epsilon_h)}.
\label{eq5}
\end{eqnarray}
\end{subequations}
Here $v=(4\pi/3)R^3$ is the volume of nanoparticles, $\epsilon$ ($\epsilon_h$) is the dielectric function of the nanoparticle material (background medium), and $G_\circ=i(k^3/6\pi)$.
By using a linear complex vector space ${\mathrm C}^{3N}$, Eq.~(\ref{eq3}) can be written in a more compact notation
\begin{equation}
|{\bf P}\rangle=|{\bf P}^i\rangle+|{\bf P}^f\rangle
\label{eq6}
\end{equation}

Where $|{\bf P}\rangle=({\bf P}_1,{\bf P}_2,\cdots,{\bf P}_N)$ representing 3N-dimensional vector of dipole moments. As an arbitrary vector in this space, $|{\bf U}\rangle$ denotes a column vector and $\langle{\bf U}|$ denotes a row vector with the complex conjugated elements. More over, $\langle{\bf \bar U}|$ represents a row vector with exactly the same elements as $|{\bf U}\rangle$, and $|\bar{\bf U}\rangle$ represents a column vector like $|{\bf U}\rangle$ with complex conjugated elements. The Cartesian components of the $i$-th individual of an arbitrary vector $|{\bf U}\rangle$, can be expresses as $U_{i\alpha}=\langle i\alpha|{\bf U}\rangle$. Here, the Greek indexes stand for the cartesian components (i.e., $\alpha,\beta=x,y,z$) and $\langle\cdot|\cdot\rangle$ denotes the standard inner product on ${\mathrm C}^{3N}$. Moreover an orthogonal standard basis set $\{|i\alpha\rangle\}$ can be defined as 
\begin{equation}
\{|i\alpha\rangle\}={\Bigg\{}
\begin{bmatrix} 1\\0\\ \vdots \\0\end{bmatrix},
\begin{bmatrix} 0\\1\\ \vdots \\0\end{bmatrix},\cdots,
\begin{bmatrix} 0\\0\\ \vdots \\1\end{bmatrix}{\Bigg\}},
\label{eq7}
\end{equation}
with property
\begin{equation}
\langle i\alpha|j\beta\rangle=\delta_{ij}\delta_{\alpha\beta}~(i,j=1,2,\cdots,N).
\label{eq8}
\end{equation}
Re-writing Eq.~(\ref{eq4}) in the introduced complex vector/matrix space ${\mathrm C}^{3N}$, and inserting in Eq.~(\ref{eq6}) gives
\begin{equation}
Z|{\bf P}\rangle- \hat{\mathbb W} |{\bf P}\rangle=Z|{\bf P}^f\rangle
\label{eq9}
\end{equation}
where ${\hat {\mathbb W}}$ is a $3N\times 3N$ block matrix, representing the dipolar interaction between nanoparticle. This complex-symmetric matrix, namely the {\it interaction matrix}, is:
\begin{equation}
\hat{\mathbb W}=
\begin{bmatrix}
{\hat{\bf 0}} & {\hat{\bf G}}_{12} & \cdots & {\hat{\bf G}}_{1N}\\ {\hat{\bf G}}_{21} & {\hat{\bf 0}}& \cdots & {\hat{\bf G}}_{2N}\\ \vdots& \vdots & \ddots & \vdots\\{\hat{\bf G}}_{N1} & {\hat{\bf G}}_{N2}& \cdots &{\hat{\bf 0}}
\end{bmatrix}
\label{eq10}
\end{equation}
and we adopted the spectral variable \cite{ZZmarkel}
\begin{equation}
Z(\omega)\equiv1/\alpha(\omega)=-[X(\omega)+i\delta(\omega)].
\label{eq11}
\end{equation}
Using Eq.~(\ref{eq5}), we obtain
\begin{subequations}
\begin{eqnarray}
X(\omega)&=&-\frac{v^{-1}}{3}\Big(1+3\epsilon_h\frac{\epsilon'-\epsilon_h}{|\epsilon-\epsilon_h|^2}\Big)~,\\
\delta(\omega)&=&v^{-1}\frac{\epsilon_h\epsilon''}{|\epsilon-\epsilon_h|^2}+{\tt Im}G_\circ.
\end{eqnarray}
\end{subequations}
The variable $X(\omega)$ can be used as a frequency parameter which shows the proximity of $\omega$ to the resonance frequency of nanoparticles. For special case of spherical nanoparticles, $X\sim 0$ occures for $\epsilon\sim -2\epsilon_h$. The variable $\delta(\omega)$ is the dielectric losses and can be used to calculate the resonance quality factor which is proportional to $\delta^{-1}$.

The interaction matrix is symmetric ${\mathbb W}_{ij}={\mathbb W}_{ji}$, and the blocks are the dyadic green's ${\mathbb W}_{ij}=\hat{\bf G}_{ij}$, which are $3\times 3$ complex symmetric matrix. As we should know by now, while the specific expressions for the interaction matrix elements are basis-dependent, the symmetry properties of the matrix and the blocks are basis-independent. 
Because $\hat{\mathbb W}$ is complex-symmetric, we have $\hat{\mathbb W}\neq \hat{\mathbb W}^\dag$ which implies that the eigenvectors are not orthogonal in general. However, it can be shown that the eigenvectors are linearly independent. This is in accordance with the special case of quasistatic approximation where the interaction matrix is purely real and so is Hermitian. Suppose $w_n$ is an eigenvalue of the interaction matrix $\hat{\mathbb W}$, and $|n\rangle$ is a corresponding normalized eigenvector in a linear complex vector space ${\mathrm C}^{3N}$. We have
\begin{equation}
\hat{\mathbb W} |n\rangle=w_n|n\rangle~~~~~n=1,2,\cdots,3N
\label{eq12}
\end{equation}
where $w_n$ is complex in general. We denote the eigenvalues spectrum by a nonempty finite set $\sigma(\hat{\mathbb W})$, which at most contain $3N$ distinct element and from Eq.~(\ref{eq10}), we have ${\tt Tr}(\hat{\mathbb W})=\sum_n w_n=0$. The spectral radius of $\hat{\mathbb W}$ defines by $\rho(\hat{\mathbb W})=\max\{|w_n|:~ w_n\in\sigma(\hat{\mathbb W})\}$ which implies that every eigenvalues in a set $\sigma(\hat{\mathbb W})$ lies in the closed bounded disk $\{z\in{\mathrm C}: |z|\leq \rho(\hat{\mathbb W})\}$ in the complex plain. The spectrum and spectral radius of the interaction matrix are essential features which are independent of the choice of basis. While these features sensitively depend on the geometrical arrangement of nanoparticles, they do not depend on the nanoparticle composition material. 
An important issue to be noted is that, the complex-symmetry of the interaction matrix is a purely algebraic property, and has no affect on the spectrum of the matrix by it self. However, any regulation in the nanoparticle arrangement may results in a block \emph{ structured} interaction matrix. By a block \emph{structured} interaction matrix, we typically mean an interaction matrix whose blocks have formulaic relationship, regulation and similarity. Such regularities in the blocks, directly influences the eigenvalues and might be accompanied by degeneracy, scaling behavior and regulation in the spectrum which might also affects the spectral radius and spectrum bound. On the other side, the eigenvectors of interaction matrix are linearly independent and form a complete vector space. From the biorthogonality principle for complex-symmetric matrix we have \cite{hornmatrix}
\begin{equation}
\langle \bar m|n\rangle=0~~~~{\text if}~~~~ m\neq n,\label{eq13}
\end{equation}
where $\langle\bar m|$ is a left eigenvector associated with an eigenvalue $w_m$ of $\hat{\mathbb W}$ such that $\langle\bar m|\hat{\mathbb W}=\langle\bar m|w_m$.
This allows introducing the identity operator in terms of the interaction matrix eigenvectors as
\begin{equation}
\mathbbm{1}=\sum_{n=1}^{3N}\frac{|n\rangle\langle\bar n|}{\langle \bar n|n\rangle}.
\label{eq14}
\end{equation}
We assume that the eigenvectors are normalized, i.e., $\langle n|n\rangle$=1, however, $\langle \bar n|n\rangle$ in general, is a complex quantity. 
One should not conceive left eigenvectors as merely a parallel theoretical alternative to right eigenvectors. Each type of eigenvector can supply different information about the interaction matrix. Using the eigenvectors of the interaction matrix as a basis, Eq.~(\ref{eq9}) can be solved to calculate the dipole moments in term of fluctuating dipoles:
\begin{equation}
|{\bf P}\rangle=\sum_{n=1}^{3N}\frac{Z}{Z-w_n}\frac{|n\rangle\langle\bar n|\bf P^f\rangle}{\langle \bar n|n\rangle}.
\label{eq15}
\end{equation}
where the summation runs over all eigen-pairs $\{w_n,|n\rangle\}$ of the interaction matrix. Moreover, we have used the unity operator in the basis $|n\rangle$. 
This expression connects the total dipole moment at the position of each nanoparticle to the fluctuating dipoles. After multiplication on both sides with $\langle i\alpha|$, making use of the identity operator of standard basis $\mathbbm{1}=\sum_{j\beta}|j\beta\rangle\langle j\beta|$, the cartesian components of dipole moment of the $i$-th nanoparticle is related to the fluctuating dipoles according to
\begin{eqnarray}
\label{eq16}P_{i\alpha}&=&\langle i\alpha|{\bf P}\rangle\\ 
\notag \nonumber &=&\sum_{n=1}^{3N}\sum_{j\beta}\frac{Z}{Z-w_n}\frac{\langle i\alpha|n\rangle\langle\bar n|j\beta\rangle}{\langle \bar n|n\rangle}\langle j\beta|\bf P^f\rangle
\end{eqnarray}

Similar procedure can be used for calculating the local fields in terms of the fluctuating dipoles. To this end, we start by writing Eq.~(\ref{eq1}) in a compact notation
\begin{equation}
|{\bf E}\rangle= \hat{\mathbb G} |{\bf P}\rangle
\label{eq17}
\end{equation}
where $|{\bf E}\rangle=({\bf E}_1,{\bf E}_2,\cdots,{\bf E}_N)$ representing 3N-dimensional vector of local fields and $\hat{\mathbb G}=G_\circ \mathbbm{1}+\hat{\mathbb W}$. Since $\hat{\mathbb G}$ and $\hat{\mathbb W}$ commute it follows that $\hat{\mathbb G}|m\rangle=(G_\circ+w_m)|m\rangle$. 
In the interaction matrix basis, the solution of Eq.~(\ref{eq17}) acquires the form
\begin{equation}
|{\bf E}\rangle=\sum_{m=1}^{3N}\frac{Z(G_\circ+w_m)}{Z-w_m}\frac{|m\rangle\langle\bar m|\bf P^f\rangle}{\langle \bar m|m\rangle}.
\label{eq18}
\end{equation}
 The expression for the Cartesian components of local field at the position of {\it i}-th nanoparticle is
\begin{eqnarray}
\label{eq19}E_{i\alpha}&=&\langle i\alpha|{\bf E}\rangle\\ 
\notag \nonumber &=&\sum_{m=1}^{3N}\sum_{j'\beta'}\frac{Z(G_\circ+w_m)}{Z-w_m}\frac{\langle i\alpha|m\rangle\langle\bar m|j'\beta'\rangle}{\langle \bar m|m\rangle}\langle j'\beta'|\bf P^f\rangle.
\end{eqnarray}

The dipole moment of the $i$-th particle, ${\bf P}_i$, interacts with the local field ${\bf E}_i$ such that the total power dissipated in it is given by \cite{manybodynikbakht,many-bodyben}.
\begin{equation}
\mathcal{P}_i=\overline{[{\bf E}_i^*(t)\cdot{\dot{\bf P}}_i(t)]}=
2\int_0^\infty\omega\frac{d\omega}{4\pi^2}{\tt Im}\overline{[ {\bf E}_i^*(\omega)\cdot{{\bf P}}_i(\omega)]}.~~~~
\label{eq20}
\end{equation}
where $\overline{[\cdots]}$ represents the ensemble average. In the basis of the interaction matrix, this average becomes
\begin{equation}
{\tt Im}\big[\overline{{\bf E}_i^*(\omega)\cdot{{\bf P}}_i(\omega)}\big]={\tt Im}\sum_\alpha \overline{\langle{\bf E}| i\alpha\rangle \langle i\alpha|{\bf P}\rangle}.
\label{eq21}
\end{equation}
Incerting Eq.~(\ref{eq16}) and (\ref{eq19}) into Eq.~(\ref{eq20}) and using Eq.~(\ref{eq21}), the power dissipated in $i$-th nanoparticle can be split into two parts as
\begin{equation}
\mathcal{P}_i=\mathcal{F}_{ii}+\sum_{j\neq i}\mathcal{F}_{ij},
\label{eq22}
\end{equation}
where $\mathcal{F}_{ii}$ is associated with the power that lost by $i$-th nanoparticle due to radiation and $\mathcal{F}_{ij}$ is the power it gains due to radiation of {\it j}-th nanoparticle.
\begin{subequations}
\begin{eqnarray}
\mathcal{F}_{i}&=&\int_0^\infty\frac{d\omega}{2\pi}{\mathcal T}_{ii}(\omega)\Theta(\omega,T_i),\\
\mathcal{F}_{ij}&=&\int_0^\infty\frac{d\omega}{2\pi}{\mathcal T}_{ij}(\omega)\Theta(\omega,T_j).
\end{eqnarray}
\label{eq23}
\end{subequations}

Here, $\Theta(\omega,T)$ is the mean energy of Planck oscillator at frequency $\omega$ and at the temperature $T$ and
\begin{subequations}
\label{eq24}
\begin{eqnarray}
\label{eq24a}{\mathcal T}_{ii}(\omega)&=&4|Z|^2{\tt Im}(\chi)\Bigg[{\tt Im}(\chi)\sum_{\alpha\beta}|f_{ii}(\alpha,\beta)|^2
\\\notag\nonumber&&~~~~~~~~~~~~~~~~~~~~-{\tt Im}\sum_{\alpha}\frac{f_{ii}(\alpha,\alpha)}{Z}\Bigg],\\
\label{eq24b}{\mathcal T}_{ij}(\omega)&=&4|Z|^2\big[{\tt Im}(\chi)\big]^2\sum_{\alpha\beta}|f_{ij}(\alpha,\beta)|^2.
\end{eqnarray}
\end{subequations}
Here, ${\mathcal T}_{ii}$ and ${\mathcal T}_{ij}~(={\mathcal T}_{ji})$ are the monochromatic cooling and transmission coefficients, respectively (see Appendix for details). Moreover, we have defined a $3N$~by~$3N$ block matrix 
\begin{equation}
\label{eq25}
\hat{\bm f}= Z(Z-\hat{\mathbb W})^{-1}
\end{equation}
with elemints
\begin{equation}
\label{eq26}
f_{ij}(\alpha,\beta)=\langle i\alpha|\hat{\bm f}|j\beta\rangle=\sum_{l=1}^{3N}\frac{Z}{(Z-w_l)}\frac{\langle i\alpha|l\rangle\langle\bar l| j\beta\rangle}{\langle \bar l|l\rangle}.
\end{equation}
which is completely symmetrical, i.e., ${\bm f}_{ij}={\bm f}_{ji}$ and $f_{ij}(\alpha,\beta)=f_{ij}(\beta,\alpha)$. Moreover, $w_l$ and $|l\rangle$ are eigen pairs of the interaction matrix $\hat{\mathbb W}$.

Equations (\ref{eq24}) and (\ref{eq26}) allow us to express the cooling coefficients and transmission coefficients in terms of the eigenfunctions and eigen-frequencies of the interaction matrix. The cooling coefficients then can be inserted in Eq.~(\ref{eq23}a) to calculate the radiative cooling of each particle and summation over all particles gives the cooling rate of the structure. On the other side, the transmission coefficients can be used to calculate the net power exchanged between each pair of particles (say $i$ and $j$) in the system from the expression
\begin{equation}
\mathcal{H}_{ij}={\Big|}\mathcal{F}_{ij}-\mathcal{F}_{ji}{\Big|}.
\label{eq27}
\end{equation}
With the use of Eq.~(\ref{eq23}), we can introduce the self-conductance
$\mathcal{G}_i(T)\equiv\frac{\partial\mathcal{F}_i}{\partial T}$ and mutual-conductance $\mathcal{G}_{ij}(T)\equiv\frac{\partial\mathcal{F}_{ij}}{\partial T}$ at temperature $T$. While the mutual-conductance represents the rate at which heat flow between particles for small perturbation in temperatures, the self-conductance represents the rate of radiative cooling. All these equations are general, in the sense that they place no restriction on the geometrical arrangement of particles. In order to apply this formalism to ensemble of particles, we require that $R$ be smaller than both the thermal wavelength $\lambda_T=c\hbar/(K_BT)$ and the intra-particle distances $d$ to substantiate the dipole approximation.

Equations (\ref{eq24a}) and (\ref{eq24b}) contain complete information on the transmission and cooling coefficients dependence on the nanoparticle characteristics and geometric configuration. It is clear from Eq.~(\ref{eq24}) that the imaginary part of the susceptibility tensor of particles (corresponding to their absorption) are presented in both the cooling and transmission coefficients. In the case of composite structures we expect that the polarizability mismatches influences the heat flow. But, in case where all nanoparticles are the same, the arrangement of nanoparticles plays an important role. The cooling and transmission coefficients can show resonance due to the $Z-w_l$ term appeared in the denominator of Eq. (\ref{eq26}). The radiative properties of the system would have features arising from these resonances, such as localization or de-localization of these modes over the structure. These modes will came into resonance whenever the denominator of $f$ tends to zero, which results in heat flux and radiative cooling enhancement. The resonant frequencies (eigen modes) satisfy $Z-w_l\rightarrow 0$, where $w_l$'s are the eigen-functions of the interaction matrix and $Z$ is the spectral variable.  For an arbitrary collection of $N$ interacting nanoparticles with volume $v=(4\pi/3)R^3$, there exist at most $3N$ number of such  modes that contribute to the resultant heat exchange between particles (and also radiative cooling of particles) in the system. The resonance frequencies of these modes can be tuned by varying arrangement, size, and composition of nanoparticles. 

While the resonance frequencies of these modes depend on the arrangement of particles through $w_l$'s, the influences of size and material composition are accounted by the spectral variable $Z=Z(R,\omega)$. It is also clear from Eq.~(\ref{eq26}), that the presence of an additional particle to the system, adds $3$ number of such modes that alters the heat transfer property and particles radiative cooling in the system. Moreover, for a given distribution of nanoparticles, the resonance frequencies of these modes are identical for all particles cooling rates and also the heat flux between each pair of particles in the system. Due to the structural characteristics, both the modes and their weights depend sensitively on the presence or absence of any symmetry in the structure like transnational or scaling properties.

\begin{figure}
\centering
\includegraphics[]{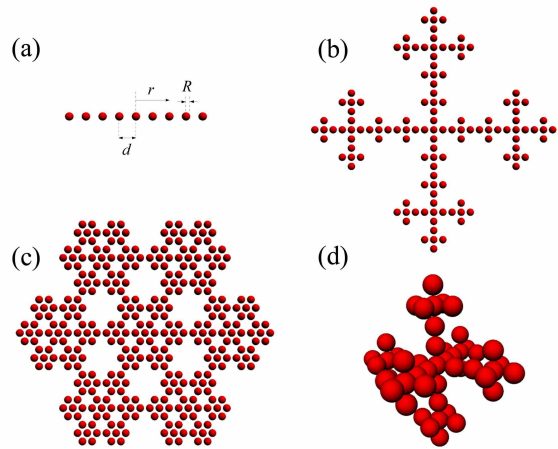}
\caption{ Schematic illustration of Vicsek fractals (composed of the same nanoparticles, say, $N$ sphere of the same radius $R=5$~nm) of  functionality $F$ (i. e., number of nearest neighbors of the branching sites): (a) $F=2$ (VF2); (b) $F=4$ (VF4); (c) two-dimensional Vicsek fractal with $F=6$ (VF6) and (d) three-dimensional Vicsek fractal with $F=6$ (3D-VF6). The growing fractals formed  by  repeated  addition  of  copies  of  the initial configuration of nanoparticles. Moreover, $d=3R$ is the separation between nearest neighbors and $r$ is the typical separation between each pair of particles.  }\label{fig1}
\end{figure}
\section{Fractal Structures}\label{sec3}
In the previous section we introduced the formalism of radiative heat transfer in a many-body system. From equations~(\ref{eq24}) and (\ref{eq26}), it follows that, the radiative properties in a collection of $N$ nanoparticles can be described in terms of the excitation modes. The contribution of each of these modes to the heat exchange between nanoparticles or radiative cooling/heating of structure depends on the structure characteristics.  The volume filling fraction $p$ is a common way of expressing the concentration of nanoparticles in a systems. However, as discussed earlier, the collective properties of heat transfer in many-body systems strongly depend on the spatial arrangement of nanoparticles in system. As maintained earlier, the fractal dimension is another measure of spatial arrangement of nanoparticles in a structure which can be applied to both fractal and non-fractal structures.   

A fractal structure can be build by arranging nanoparticles in one, two or three dimensional Euclidean space that displays self-similarity on all scales. The self similarity (scaling behavior) is the results of the simple strategy for an initial configuration of nanoparticles that repeated over and over in an ongoing feedback loop. For the sake of simplicity we choose a Vicsek fractal \cite{vicsekfractal} with different functionality $F$ as a typical fractals as depicted in figure~(\ref{fig1}). Similar to two-dimensional (three-dimensional) fractals, Vicsek fractals have zero area (volume). These fractals made of $N$ identical nanoparticles with radius $R$.  The nanoparticles nearest separation distance in the initial configuration of each fractal is $d$. The functionality of each fractal, $F$, defines the number of nearest neighbors in the branching site of the middle particle in the initial configuration. Moreover, the separation between arbitrary pairs of particles in a fractal is denoted by $r$. As we can see the same {\it type} of pattern appears on all scales for each fractal. The number of nanoparticles (i.e, the size) in a fractal aggregate scales  with the radius of  gyration, $R_g$, as follow \cite{vicsek1992fractal}
\begin{equation}
N\sim R_g^{D_f},
\label{eq28}
\end{equation}
where $D_f$ is the fractal dimension and, in general, is a non-integer value and less than the dimension of the embedding space $D$, i.e. ($D_f<D$). The radius of gyration (for a given fractal of size $N$) is the average distance between points (nanoparticles) of the fractal and its center of mass:
\begin{equation}
R_g=\frac{1}{N}\sum_{i=1}^N|{\bm r}_i-{\bm R_{CM}}|.
\label{eq29}
\end{equation}
Here, ${\bm r}_i$ is the position of the $i$-th nanoparticle in the structure, and ${\bm R_{CM}}$ is the center of mass coordinate, where the total mass of the structure is supposed to be concentrated.
\begin{figure}
\centering
\includegraphics[height=6cm,width=6cm]{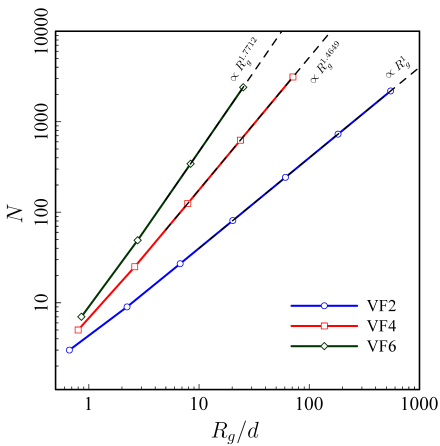}
\caption{ Double logarithmic plot of the radius
of gyration for the Vicsek fractals of $N$ particles with functionality: $F=2$ (VF2); $F=4$ (VF4); $F=6$ (VF6 for both 2D-VF6 and 3D-VF6). The dashed lines represent the asymptotic behavior of curves for large fractal sizes  ($N>100$) and the scaling exponents. The fractal dimensions are $D_f=1, 1.4649, 1.77$ for $F=2,4,6$, respectively. }
\label{fig2}
\end{figure}
Figure~(\ref{fig2}) show a log-log  plot  of $N$ versus $R_g/d$  for Vicsek fractals, where $d$ is a typical smallest separation between neighbor particles. In accordance with our definition, the  slope  of  the  fitted  line  yields the fractal dimension $D_f$ for each fractal. For the special case of Vicsek fractals, the functionality, $F$, determines $D_f$ through $D_f=\ln (F+1)/\ln 3$. As expected, the structure with $F=2$, which corresponds to a linear chain of nanoparticles, results in an integer value for fractal dimension (and hence non-fractal structure) with $D_f=D=1$. For the case of $F=4$ (VF4) the fractal dimension is increased which implies that the average number of neighbors increased for any particle in the fractal in comparison to VF2. It is clear
from the figure that the data for Vicsek fractals collapse into a single curve for both 2D-VF6 and 3D-VF6 fractals, showing that the fractal dimension for VF6 embedded in two and three dimension (2D-VF6 and 3D-VF6 respectively) are essentially the same, i.e., $D_f=1.77$. It should be emphasizes that, while 2D-VF6 and 3D-VF6 have the same fractal dimension, the geometrical arrangement of nanoparticles are totally different in these fractals. In order to distinguish between these structures one can calculate the density-density correlation function. The probability of finding two particles in the structure with separation distance $r$, is proportional to the density-density correlation function, $C(r)$, which has the power law dependency on $r$ as \cite{vicsek1992fractal}
\begin{equation}C(r)\sim r^{\gamma},
\label{eq30}
\end{equation}
with $\gamma=D_f-D$. This probability becomes constant for periodic arrangements of nanoparticles (e.g., the VF2 fractal) or even in random distribution of nanoparticles. On the other side, the density-density correlation function is a rapidly decreasing function of $r$ for fractal structures. It is clear that the probability of finding two particles in 3D-VF6 as a function of separation $r$ decreases much faster than that in 2D-VF6 fractal. We also note from Eq.~(\ref{eq28}) that $p\rightarrow 0$ for sufficiently large fractals, i.e., $R_g\rightarrow \infty$. However, there is a high probability of finding a particle in a close vicinity of any given particle in a fractal. This is obvious, since from Eq.~(\ref{eq30}) the pair correlation function is large for small distances $r$ in fractals. In the case of random distribution of particles (RGP), $p$ is very small as in fractals, however, the pair correlation function is distance independent.  

\begin{figure}
\centering
\includegraphics[height=7cm,width=8.5cm]{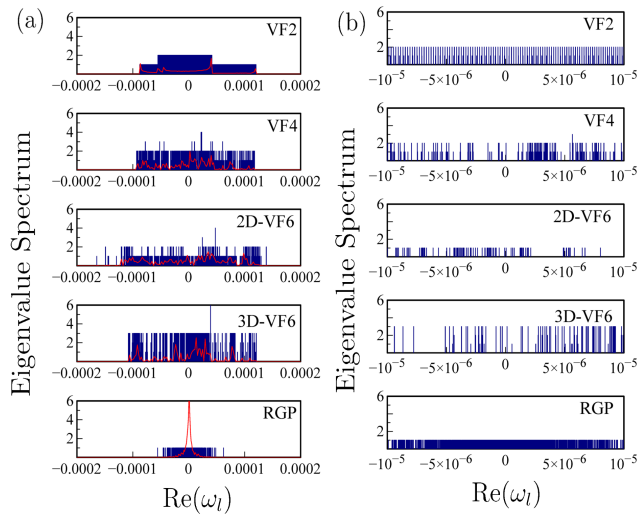}
\caption{Eigenvalue histogram (a) and small piece of the histogram (b) of the interaction matrices $\mathbb W$ for Vicsek fractals (VF2, VF4, VF6) and random gas of particles (RGP) consisting $N=1000$ nanoparticles with $d=3R$. Blue vertical lines denoting eigenvalues and red curves represent the eigenvalue density distribution for each configuration.}
\label{fig3}
\end{figure}
\section{heat flux in fractal and non-fractal structures}\label{sec4}
In Sec~(\ref{sec2}), we introduced a theoretical approach to describing the radiative heat transfer in an arbitrary ensemble of nanoparticles. Now, we are ready to apply them to fractal structures. 
The distribution of the real parts of the interaction matrix eigenvalues are shown in
Fig.~(\ref{fig3}) for Vicsek fractals introduced in Sec~(\ref{sec3}). The calculations were performed for structures consisting $N=1000$ nanoparticles with $d=3R=15$~nm and eigenvalues of the interaction matrix are calculated for a typical frequency $\omega=58\times 10^{14}$~rad/s. The whole part of the eigenvalue distributions are shown by blue vertical lines in Fig.~(\ref{fig3}a). It can be seen that the eigenvalues of the interaction matrix filling up the range $\sim \pm10^{-4}$ around $\omega'_l=0$ and both the distribution and order of degeneracy of eigenvalues depend on the morphology of the structure. It is clear from this figure that  the non-degenerate part of the spectrum approximately distributes uniformly in this range for a linear chain of nanoparticles (VF2). However, the two-fold degenerate eigenvalues are confined to a narrower bound around $\omega'_l=0$. The red curve in this figure shows the density of eigen modes (or the eigenvalue density distribution $n(w_l)$). Depending on the nanoparticle arrangement, the structure may support localized or delocalized modes. In the case of linear chain of nanoparticles, the distribution shows sharp peaks at certain characteristic values of $w_l$ and modes are delocalized, i.e., they are spread over the whole structure. 

The eigenvalues of fractal structures (VF4, 2D-VF6, or 3D-VF6) seem to be filling up this range not uniformly and broadened for higher order degenerate modes. The striking feature is that the spectra look completely different from that of VF2. It is clear that there exist spectral windows for fractal structures which we expect to influence the resonance frequencies of transmission (and cooling) coefficients. We should remark that the number of distinct eigenvalues of the interaction matrix is still much smaller than $3N$ (as in VF2) due to the symmetric nature of the fractal structures we are considering. This degeneracy can easily be wiped out by small displacement in nanoparticle positions to form a random fractal. Moreover, the scaling property in these structures, which induced several localized modes in the eigenvalue distribution, leads to reduction in the radiative heat transfer in fractals compared with periodic structures.
In order to compare the results with that of the random distribution of particles, the eigenvalue distribution of a random gas of particles (RGP) are shown in the last row in figure~(\ref{fig3}a). The RGP was generated by random distribution of particles in a spherical volume with the same volume fraction of 3D-VF6 structure. the spectrum shows remarkable symmetries and localization around $\omega'_l=0$. One can see that the eigenvalue density distribution in this case is localized to a narrow range which  is much smaller than that of fractal structures. Figure~(\ref{fig3}b) shows a small piece of the eigenvalue spectrum. It can be seen that the eigenvalue statistics extended to even smaller scales. However, the eigenvalues always localized around $\omega'_l=0$ in the case of RGP.

While any symmetry in geometric arrangement of nanoparticles (e.g., periodic arrangement or deterministic on-lattice fractals) my accompanied by increasing the number of degenerate eigen-modes, the number of {\it distinct} modes  involved in the heat exchange and radiative cooling are depending sensitively on separation distance $d$ and frequency $\omega$. The width of the spectrum decreases by increasing $d$ due to weak coupling between particles at large distances, however, it does not depend sensitively on $\omega$. The crucial point to be noted is that the weight with which a mode contributes to the resultant transmission coefficient and cooling rates, depends on the numerator of Eq.~(\ref{eq26}), and thus on the symmetries in the eigenvectors $|l\rangle$ of the interaction matrix and the intensity of these radiative modes $\langle i\alpha|l\rangle^2$ at the position of each particle.

\subsection{Mutual conductance in fractals}\label{sec4-1}

In this section we discuss the influence of the fractality on heat transfer for ensemble of particles. We consider an ensemble made up of $N=1000$ small spherical silver nanoparticles with radius $R=5$~nm and lattice space $d=3R$. For the complex dielectric function $\epsilon(\omega)$ of silver, we used the Lorentz-Drude model \cite{Rakicepsilon}  
 \begin{equation}
\epsilon(\omega)=1-\frac{\Omega_p^2}{\omega(\omega-i\Gamma_0)}+\sum_{j=1}^k\frac{f_j\omega_j^2}{(\omega_j^2-\omega^2)+i\omega\Gamma_j}
\end{equation}

where $\Omega_p=\sqrt{f_0}\omega_p$ is the plasma frequency associated with intraband transitions with oscillator strength $f_0=0.845$ and damping constant $\Gamma_0=0.048$~eV. Moreover 
$f_1=0.065$, $\Gamma_1=3.886$~eV, $\omega_1=0.816$~eV,
$f_2=0.124$, $\Gamma_2=0.452$~eV, $\omega_2=4.481$~eV,
$f_3=0.011$, $\Gamma_3=0.065$~eV, $\omega_3=8.185$~eV,
$f_4=0.840$, $\Gamma_4=0.914$~eV, $\omega_4=9.083$~eV,
$f_5=5.646$, $\Gamma_5=2.419$~eV, $\omega_5=20.29$~eV. 
\begin{figure}
\centering
\includegraphics[height=5.25cm,width=7cm]{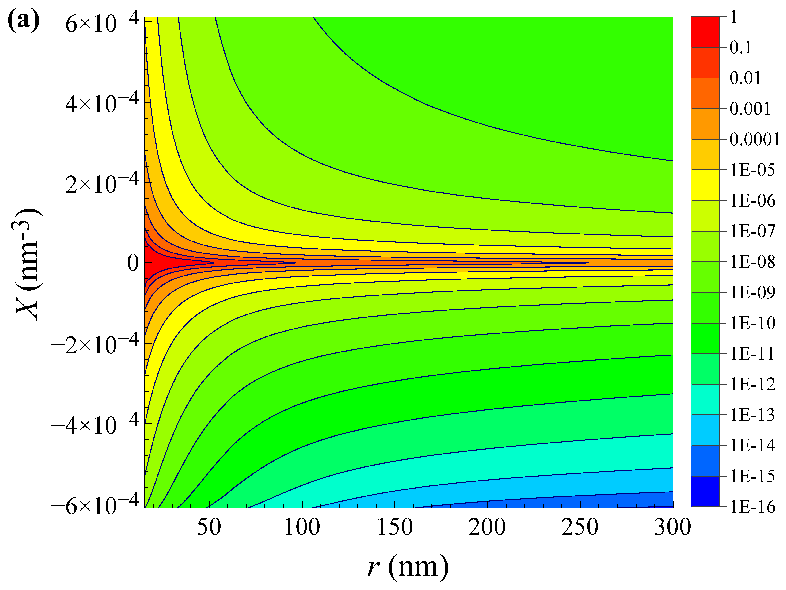}
\includegraphics[height=5.25cm,width=7cm]{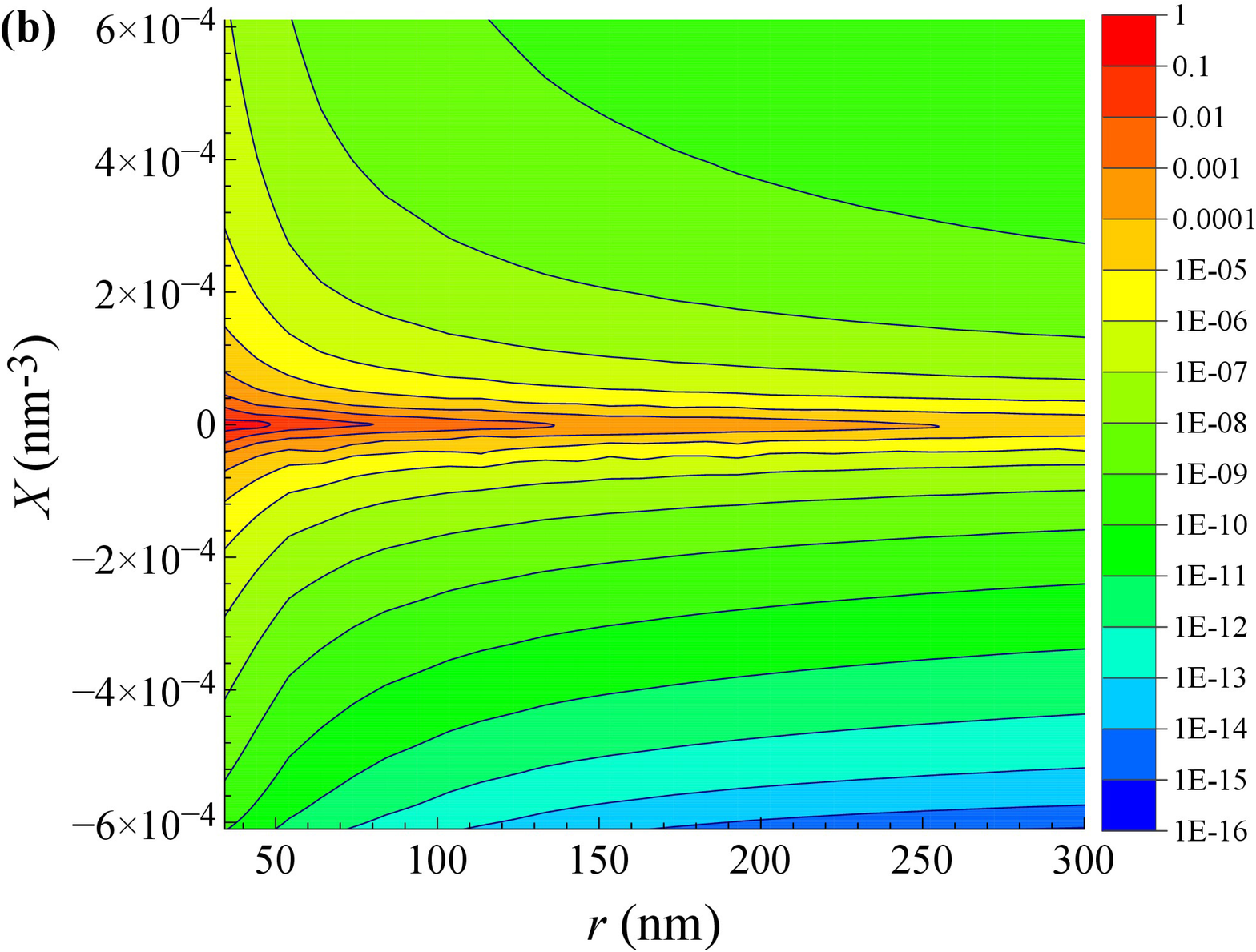}
\caption{ Transmission coefficient between two Ag nanoparticle with $R=5$~nm in (a) a two-body system, (b) inside a random gas of particles, as a function of separation distance $r$ and spectral variable $X$. }
\label{fig4}
\end{figure}

Equation~(\ref{eq24b}) is the basic equation that can be used to evaluate the transmission coefficient between each pair of particles (say $i$-th and $j$-th) in an ensemble of $N$ nanoparticles. In the case of an isolated dimmer (i.e., two-body system where $N=2$) the transmission coefficient between particles as a function of the separation distance $r$ and spectral variable $X$ is shown in Fig.~(\ref{fig4}a). It can be seen that the transmission coefficient is localized around $X=0$ where the surface plasmon resonance modes is dominant for isolated nanoparticles. Moreover, the rapid decrease in the transmission probability by increasing distance is responsible for the decreasing form of the near-field heat transfer, i.e., $\sim r^{-6}$ at small separation distances.

We expect a change in the transmission coefficient between particles when they are not isolated-pairs but placed inside a collection of $N$ nanoparticles. To assess radiative transport through the structure we evaluate the average transmission coefficient and also the average mutual-conductance over particles with same separation distance.  For this purpose, we introduced the average transmission coefficient as:
\begin{equation}
\langle {\mathcal T}(r,X) \rangle=\frac{1}{n}\sum_{i=1,j>i}^N{\mathcal T}_{ij}\delta(r_{ij}-r)H(R_c-|{\bm r}_i|).
\end{equation}
where $H(x)$ is Heaviside step function, $n=\sum_{i=1}^N \Theta(R_c-|{\bm r}_i|)$ and $R_c=R_g/4$ is a cutoff distance from the reference particle which is used to avoid boundary effects. On the other hand, the calculation of mean transmission coefficient is performed for those pair of particles in which one of their component positioned within the sphere of radius $R_c$ around the reference particles. Notice here that the ensemble average is not required for deterministic fractals. However, one may need to calculate the ensemble average of mean transmission coefficient in case of random fractals or RGP. 

The mean transmission coefficient between particles in a random gas of particles (dilute RGP) is shown in figure~(\ref{fig4}b). The striking feature is that the results look very nearly the same to that of two-body system.  The resonance frequencies in such cases congregate to the resonance of susceptibility of an individual particles (surface modes: $Z\rightarrow 0$ i.e., $X(\omega)\approx 0$), occurring for a spherical particle at $\epsilon(\omega)=-2\epsilon_h$ (where, $\epsilon_h$ and $\epsilon$ are the complex dielectric function of the background medium and particle respectively). 
\begin{figure}
\centering
\includegraphics[height=5.25cm,width=7cm]{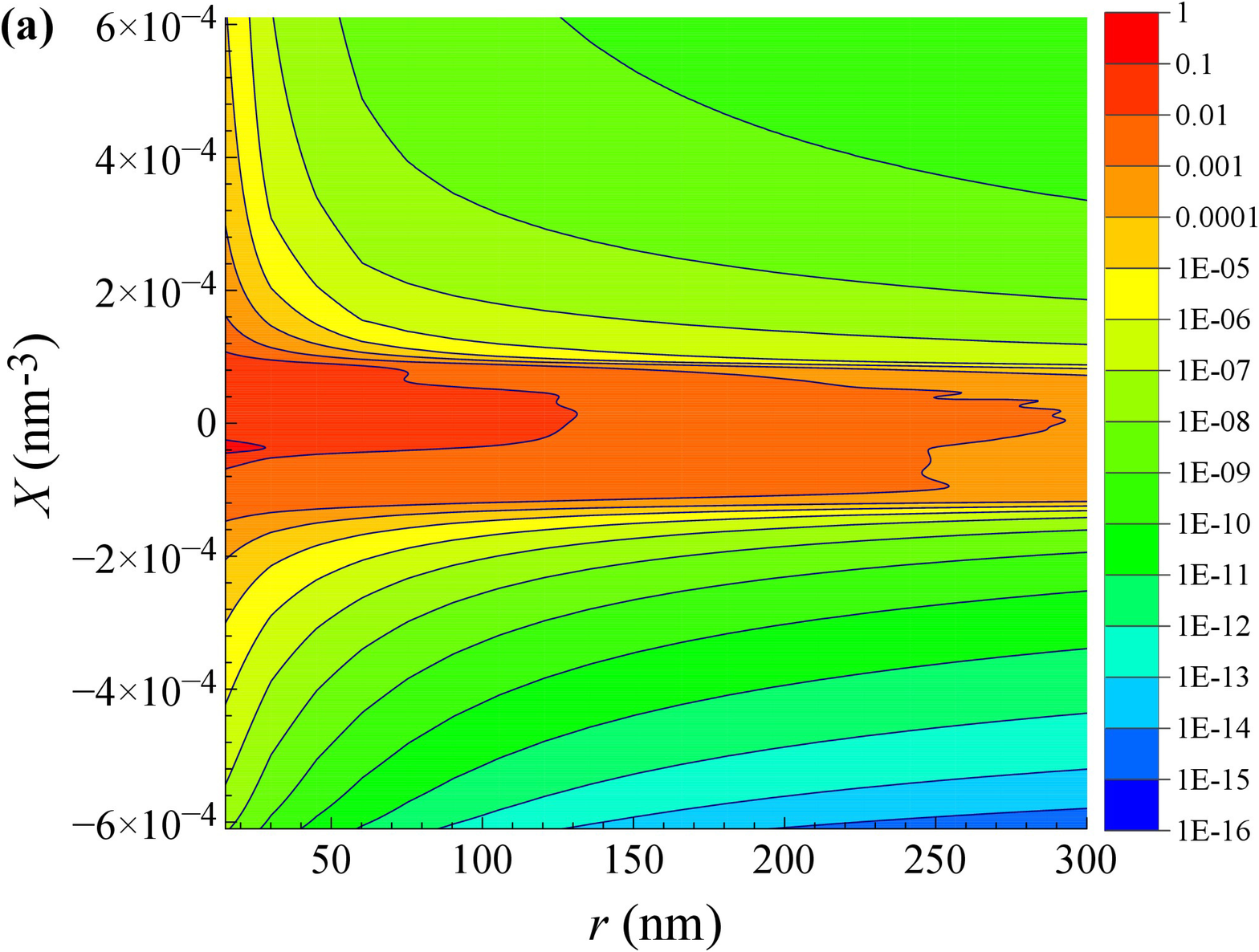}
\includegraphics[height=5.25cm,width=7cm]{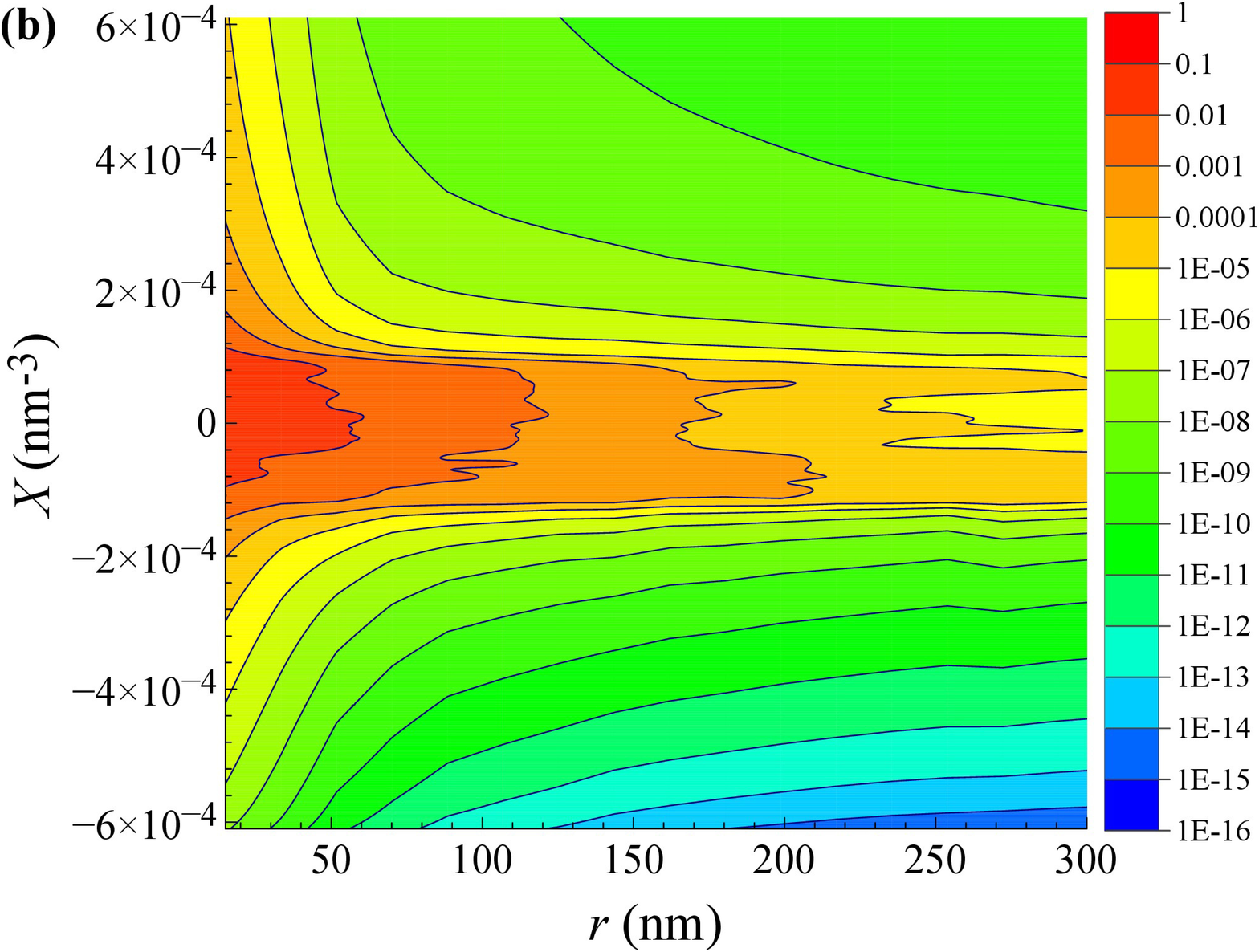}
\includegraphics[height=5.25cm,width=7cm]{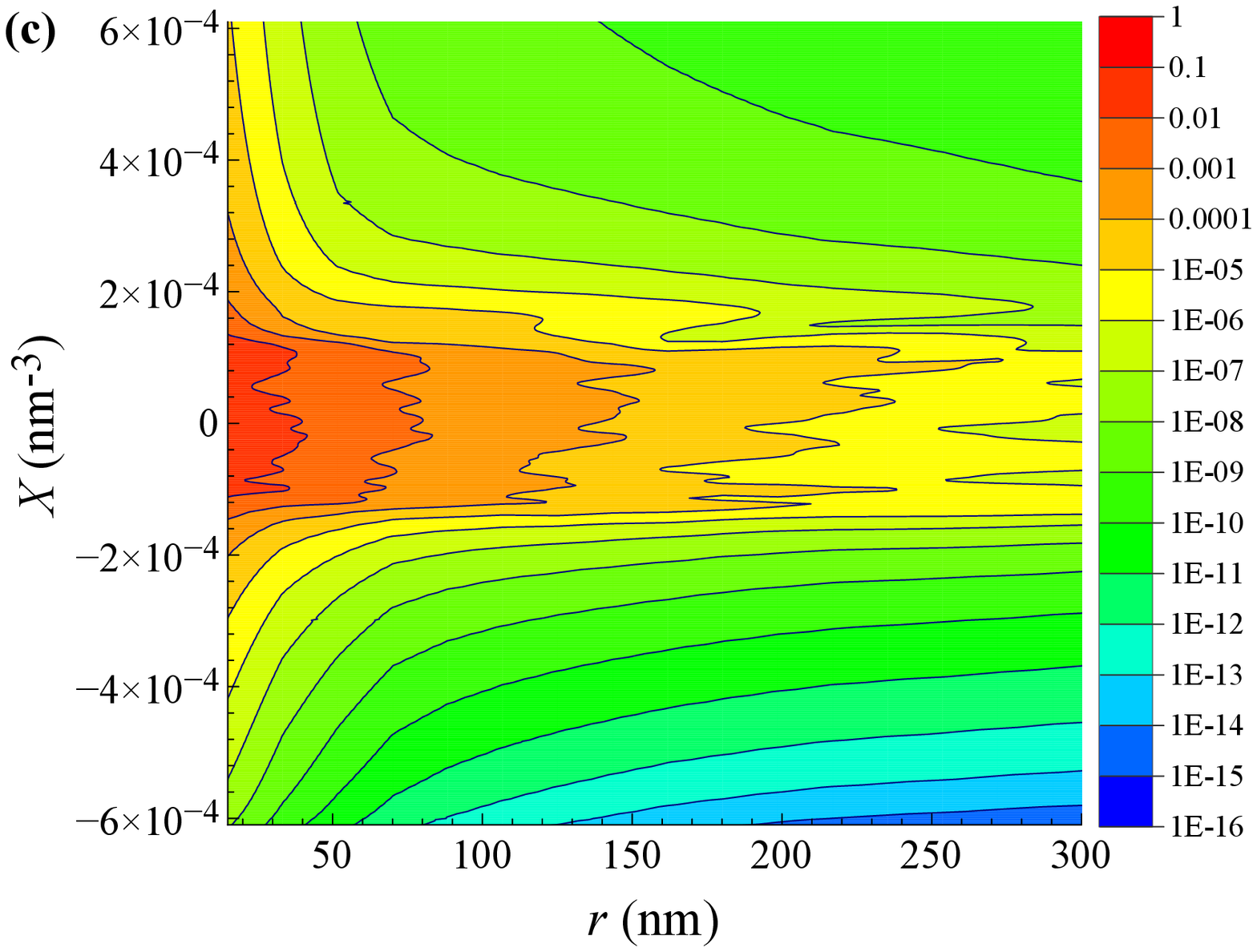}
\includegraphics[height=5.25cm,width=7cm]{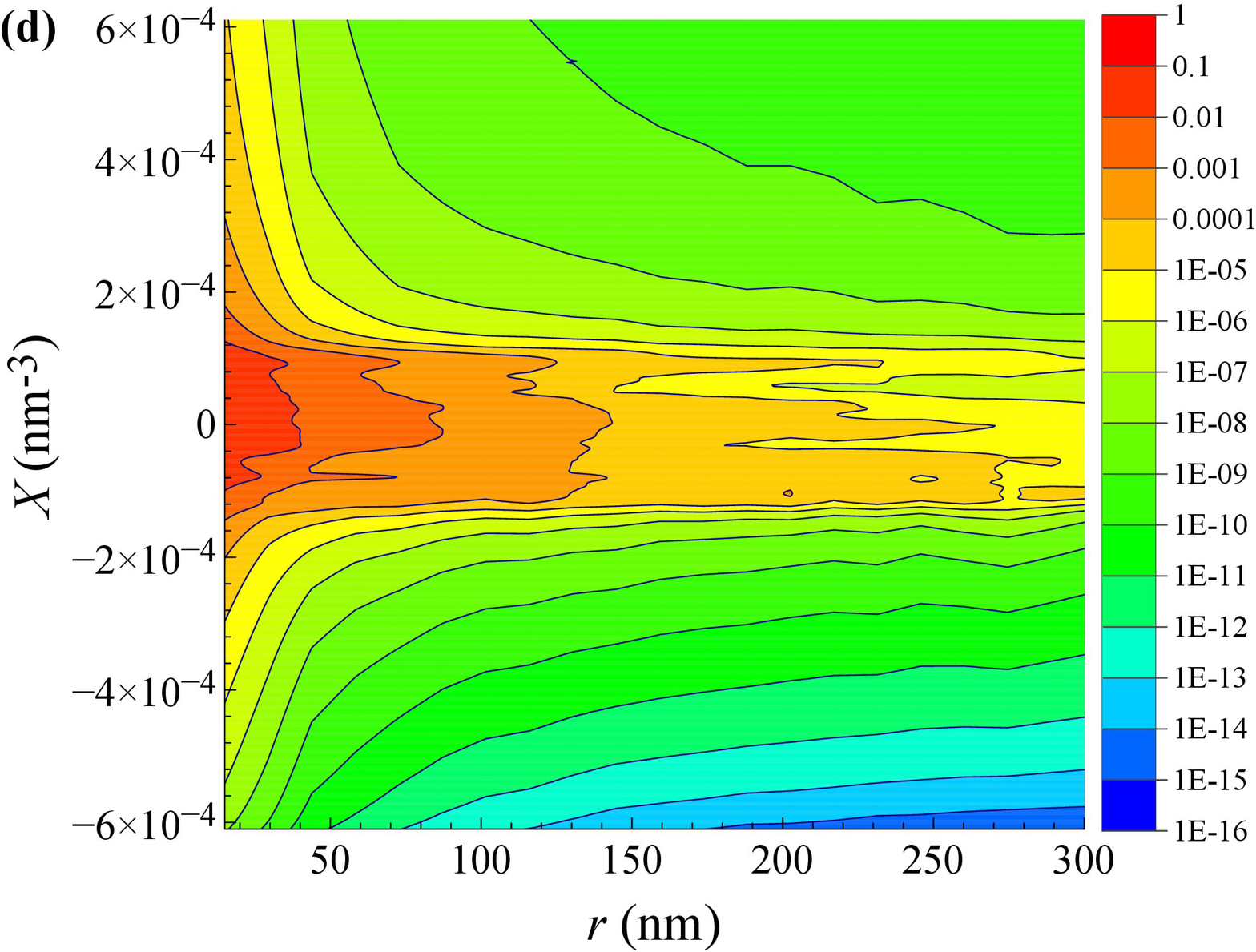}
\caption{ Color map of average transmission coefficient between pair of particles in a Vicsek fractal as a function of pair separation distance $r$ and spectral variable $X$ for (a) VF2; (b) VF4; (c) 2D-VF6 and (d) 3D-VF6.}
\label{fig5}
\end{figure}

Fig.~(\ref{fig5}) shows the average transmission coefficient between nanoparticles  inside a Vicsek fractals in terms of the separation distance $r$ and spectral variable $X$. As clearly seen in the figure, the spectral width of the transmission coefficients broadened in comparison with that of isolated two-body system. The broadening of the transmission coefficient around $X(\omega)=0$ arises from the participation of {\emph collective modes} (i.e., plasmons) in heat transport which is the special character of many-body systems. From figure~(\ref{fig5}a), it is evident that this broadening is approximately homogeneous for linear chain of nanoparticles (i.e., VF2). This feature is due to the translation symmetry in VF2 structure. On the other hand, thermal excitations $\langle i\alpha|l\rangle$ are not localized in small areas of the chain and can came into resonance simultaneously. Moreover, the transmission coefficient in a linear chain of particles decays slower as $r$ increased in comparison with two-body system. Thus, as expected, the heat transfer would be of long-range character in periodic arrangement of nanoparticles. The increase in the transmission coefficient at long-wavelength can be related to excitation of zero modes where $X(\omega)\rightarrow X_0=\frac{1}{4\pi R^3}$. When the separation between nanoparticles decrease, $d\rightarrow 2R$, heat transfer is primarily due to excitation of these modes and is large because of its resonance character. In the case of dilute RGP, there would be no transmission resonance in this parts of the spectrum (see Fig.~(\ref{fig4}b)), and consequently the collective effects is small in these structures.

\begin{figure}
\centering
\includegraphics[height=6cm,width=8cm]{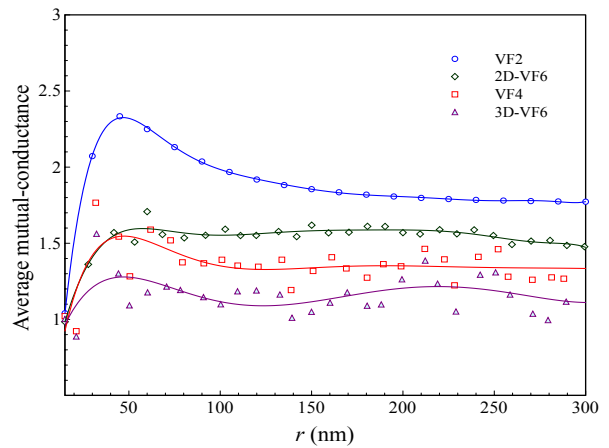}
\caption{ Average mutual conductance in collection of Ag nanoparticles at temperature $T=300$~K as a function intra-particle distance $r$. The result at each distance is normalized to the conductance between two isolate nanoparticles at same distance. Each configuration consists $N=1000$ spherical nanoparticle with $R=5$~nm.  }
\label{fig6}
\end{figure}

 The transmission coefficient in fractal structures (VF4, 2D-VF6, and 3D-VF6) are shown in figures~(\ref{fig5}b-d). The first striking feature is that the transmission spectra are broadened as in VF2. However, the broadening is inhomogeneous and red-shifted in comparison to that of VF2. This inhomogeneity is the result of local anisotropy in particle arrangements in these structures and is in agreement with frequency selective windows discussed in previous section. On the other hand, thermal excitations are localizing in small areas of fractal structures and came into resonance at different frequencies. Moreover, the transmission coefficient is rapidly decreasing function of distance in comparison to VF2. Thus, even though collective modes participate in the transmission coefficient, the heat transport is expected to be of small-range character in fractals.   
In order to illustrate this property, in Fig.~(\ref{fig6}) we show the average mutual conductance of dimmers inside the collection of particles for different structures. The calculation performed at temperature $T=300$~K and results for each distance is normalized to the thermal conductance of an isolated dimmer with same inter-particle distance. This figure shows that the mutual conductance enhanced at almost all separation distances in comparison with the two-body system. The origin of this enhancement is the existence of collective modes participating in heat transfer and is a criteria which can be used to classify the range of effective heat flux in these structures. It can be seen that the thermal conductance in VF2 is large in comparison to the other type of Vicsek fractals. It is clear, since the pair correlation function is constant in VF2, i.e., $\gamma$=0. On the other side the exponent $\gamma$ of the pair correlation function, in Eq.~(\ref{eq30}), is $-0.2288$ and $-0.5359$ for 2D-VF6 and VF4, respectively which decreases to $-1.2288$ for 3D-VF6 fractal. This indicates that the 3D-VF6 has the smallest effective range for heat transfer among these structures.  

\subsection{Radiative cooling of fractal structures}\label{sec4-2}
In this section we will investigate the influence of nanoparticle arrangement on the radiative cooling of the structure. In the absence of thermal bath, i.e., $T_b=0$, the structure self-conductance is responsible for a radiative cooling of the structure and determines how fast an ensemble of particles cools down due to radiation. This argument almost hold for case in which $T_b\neq 0$, because the magnitude of the interaction with thermal bath is much smaller than the interaction which take place inside the structure. On the other hand, the latter are in near-field while the former occur in far field regime.

We apply the collective model developed earlier to calculate the cooling coefficient of each particle in a structure. The self-conductance of a nanoparticle $\mathcal{G}_i(T)$ determines the power it losses when it is placed inside an ensemble of $N-1$ nanoparticles. It is clear that this value for the self-conductance would be different from that of isolated nanoparticle. By calculating self-conductance of all particles in a structure, we define the structure cooling conductance as
\begin{equation}
\mathcal{G}_{N}(T)=\sum_{i=1}^N\mathcal{G}_i(T)=N\langle\mathcal{G}\rangle,
\end{equation}
where $\langle\mathcal{G}\rangle$ is particles self-conductance averaged over all particles in the structure, namely, the self-conductance per particle. As mentioned earlier, $3N$ families of modes are participating in the radiative cooling of a given structure, so, the structure conductance and average self-conductance are expected to depend on the structure size $N$. 

\begin{figure}
\centering
\includegraphics[height=6cm,width=8cm]{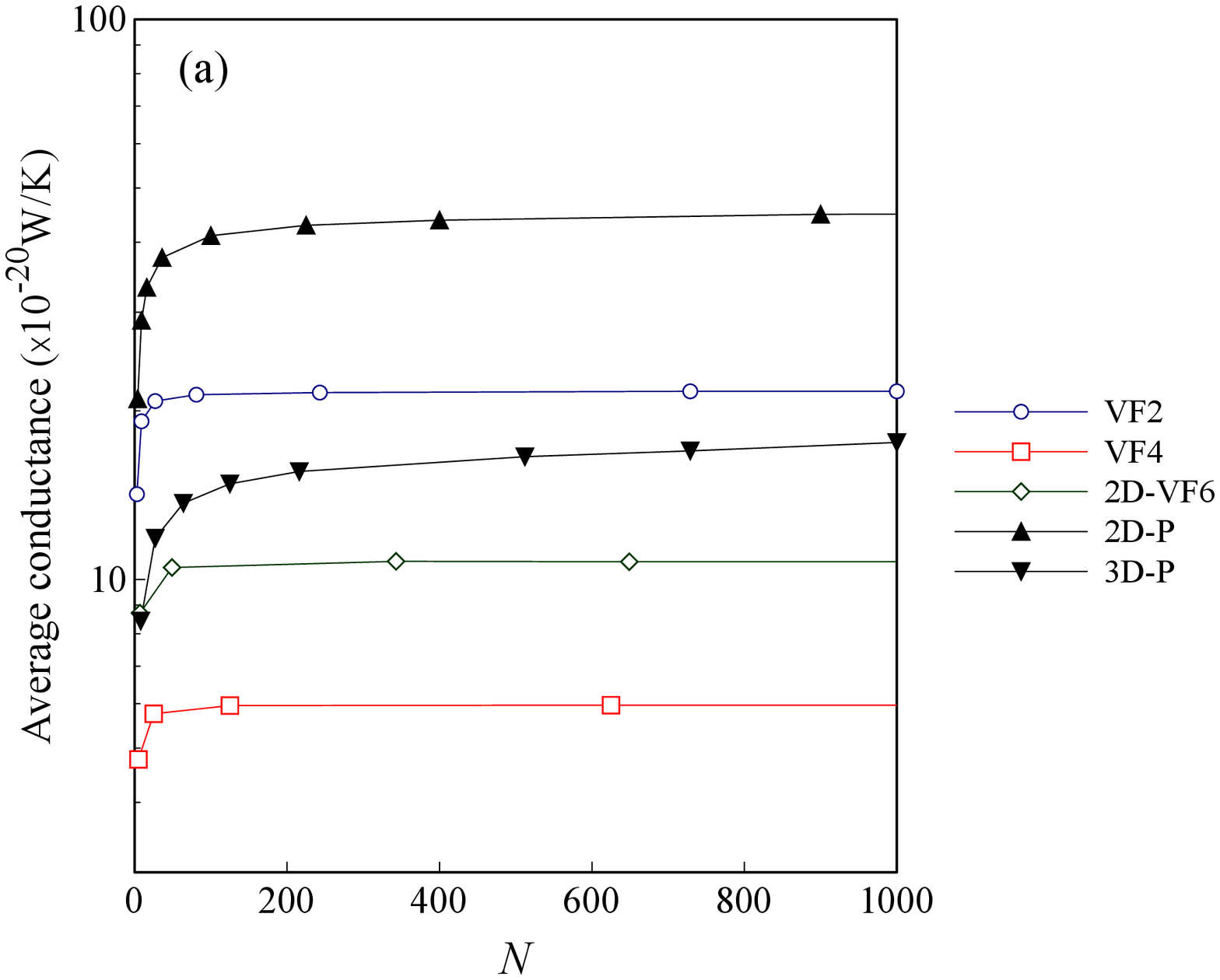}
\includegraphics[height=6cm,width=8cm]{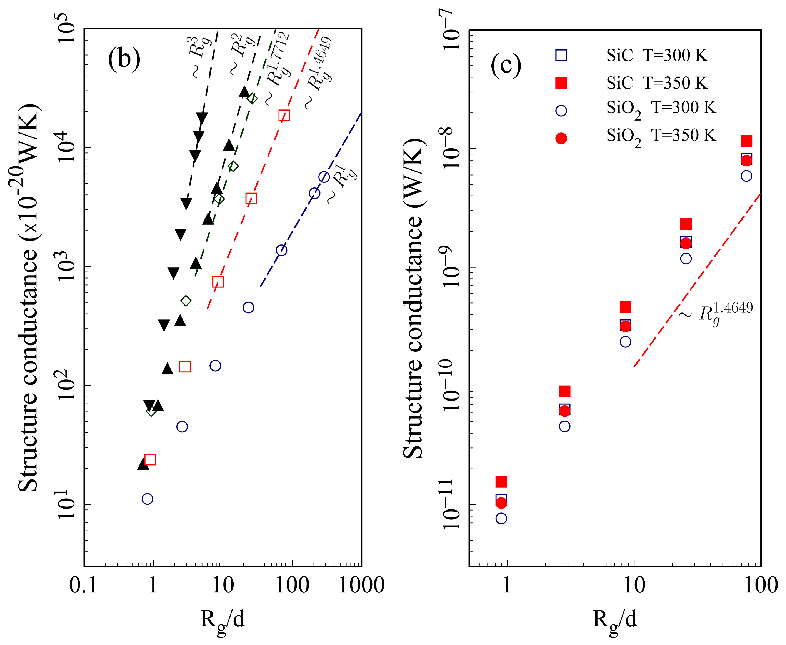}
\caption{ (a) Average conductance of particles in a structure as a function of structure size $N$ for Vicsek fractal of functionality $F=2, 4, 6$ and periodic arrangement of nanoparticles in two (2D-P) and three (3D-P) dimension, (b) The structure self conductance as a function of normalized gyration radius for both fractal and non-fractal structures. (c) Structure conductance for VF4 fractals of polar materials (SiC and SiO$_2$) at temperature $T=300$~K and $T=350$~K as a function of normalized gyration radius.}
\label{fig7}
\end{figure}

In Fig.~(\ref{fig7}a), we present an average of particles self-conductance as a size of the structure for several geometrical arrangements. It is clear that the average self-conductance enhanced by increasing the structure size and saturates for large structure sizes.  and is independent of the type of arrangement. This implies that there exists a certain characteristic length at which particles could exchange energy with each other in beyond this length, the coupling is ignitable. On the other hand, further increase in the structure size does not influence the radiative cooling of nanoparticles which can be regarded as {\emph screening effect}. The saturation of average self-conductance occurs for smaller sizes in Fractals (VF4, VF6) in comparison to periodic structures (VF2). Moreover, one notice from this figure that for a given structure size, the conductance per particle is smaller in fractal structures. This result is in agreement with the long-range character of heat flux in VF2 structure. In order to compare the results with periodic configuration at higher dimensions, the calculation performed for two-dimensional an three-dimensional periodic arrangements of nanoparticles. The 2D-P (3D-P) structure is made by periodic arrangement of nanoparticles on a cubic lattice with lattice constant $d$ in a circle (spherical) region. For 2D-P and 3D-P, the same behavior as for VF2 is observed. Once again, we notice that the saturation size is larger for 2D-P and 3D-P in comparison to fractals. This difference is not surprising, since in contrast to ordered media, the plasmon modes are localize in fractal structures and does not resonate simultaneously, which causes the average self-conductance to be smaller in fractal structures. 
The log-log plot of the structures self-conductance as a function of their gyration radius are shown in Fig.~(\ref{fig7}b). Inspection of results shows that structure total conductance increase in power law form $\sim R_g^{D_f}$. This confirms that the fractal dimension $D_f$ of the structure plays a fundamental role on radiative properties.
The dashed lines in figure.~(\ref{fig7}b) represent a power-law fit for large cluster sizes. The computed exponent is exactly the same as the fractal dimensions we have calculated from Eq.~(\ref{eq28}). As it is clear the exponent for 2D-P and 3D-P are the dimension of the embedding space of theses structures which are $D=2$ and $D=3$ respectively. It should be emphasizes that this scaling behavior is a universal properties of radiative cooling in many-body systems and does not depends on nanoparticles characteristics. To confirm this universality, we calculated the structure conductance for polar material. We used SiC and SiO$_2$ as typical materials and the calculations of structure conductance are performed at two temperatures $T=300$~K and $T=350$~K. In Fig.~(\ref{fig7}b) we only present the results of the structure conductance for VF4 fractals. It is clear from this figure that there is a power law relation between structure conductance and the radius of gyration for polar materials. Moreover, one notices that while the scaling behavior does not depend on nanoparticles characteristics, it does not depend on the structure temperature either.
\section{conclusion}\label{sec5}
 We studied the implications of the structure morphology on the radiative properties. For this purpose we proposed a new representation for radiative heat transfer formalism in many-body systems. We showed how the interaction matrix representation can explicitly feature the contribution of the nanoparticles characteristics as well as their geometric arrangement on the heat flux in many-body systems. It is shown that the heat transfer could be addressed in terms of excitation modes. We discussed the way that the strongly localized modes in fractal structures as well as the de-localized modes in periodic structures, show up in the heat transfer and radiative cooling of structures. In particular, we showed that that the radiative heat transport in highly branched fractal structures is of small range character which differs significantly from that of periodic arrangement of nanoparticles. Moreover, we showed that there exists a universal scaling behavior in structure self-conductance which holds for both fractal and non-fractal structure. 
 \section*{Acknowledgement}
 The author would like to thank Prof. Philippe Ben-Abdallah for  helpful discussions. 
 
\begin{widetext}
\appendix*
\section*{APPENDIX} 
\renewcommand{\theequation}{A.\arabic{equation}}
\setcounter{equation}{0} 
\section{Thermal modes}
To describe the thermal properties of the structure in terms of thermal excitation modes, we consider a system of $N$ identical nanoparticles of identical radius $R$. Nanoparticles are located at points ${\bm r}_i, i=1,\cdots,N$  in structure and maintained at temperatures $T_i$ inside a thermal bath at temperature $T_{b}$. 
The power dissipated in the $i$-th particle is given by
\begin{equation}
\mathcal{P}_i=\overline{[{\bf E}_i^*(t)\cdot{\dot{\bf P}}_i(t)]}=
2\int_0^\infty\omega\frac{d\omega}{4\pi^2}{\tt Im}\overline{[ {\bf E}_i^*(\omega)\cdot{{\bf P}}_i(\omega)]}.~~~~
\label{a1}
\end{equation}
Putting Eqs.~(\ref{eq16}) and (\ref{eq19}) into Eq.~(\ref{eq21}), the functional in Eq.~(\ref{a1}) would be

\begin{equation}
\overline{{\bf E}_i^*(\omega)\cdot{{\bf P}}_i(\omega)}=\sum_{mn}\sum_\alpha\sum_{j'\beta'}\sum_{j\beta}\frac{(G_\circ^*+w_m^*)Z^*Z}{(Z^*-w_m^*)(Z-w_n)}\frac{\langle i\alpha|n\rangle\langle\bar n|j\beta\rangle
\langle m| i\alpha\rangle\langle j'\beta'|\bar m\rangle}{\langle \bar n|n\rangle\langle m|\bar m\rangle}\overline{\langle j\beta|\bf P^f\rangle\langle \bf P^f|j'\beta'\rangle}
\label{a2}
\end{equation}
The last term in Eq.~(\ref{a2}) is the correlation between fluctuating dipoles and from fluctuation electrodynamics it can be written as 
\begin{equation}
\label{a3}
\overline{\langle j\beta|\bf P^f\rangle\langle \bf P^f|j'\beta'\rangle}=2\pi\hbar\delta_{jj'}\delta_{\beta\beta'}\big[1+2n(\omega,T_j)\big]{\tt Im}(\chi_j),
\end{equation}
with $n(\omega,T)=[\exp(\frac{\hbar \omega}{ k_B T})-1]^{-1}$ is the Bose-Einstein energy distribution function of a quantum oscillator at temperature T. The $\chi= \alpha+\alpha G_{0}^*\alpha^*$ is susceptibility of nanoparticle which is defined non-negatively to gives a correct direction for the heat flux between particles. Using Eq.~(\ref{eq11}), the susceptibility can be written in terms of the spectral variable $Z$ as
\begin{equation}
\label{a4}
ZZ^*\chi-Z^*=G_\circ^*.
\end{equation}
Substituting Eq.~(\ref{a3}) and (\ref{a4}) into Eq.~(\ref{a2}) and taking the imaginary part yields
\begin{equation}
\label{a5}
{\tt Im}\big[\overline{{\bf E}_i^*(\omega)\cdot{{\bf P}}_i(\omega)}\big]=\frac{4\pi}{\omega}{\tt Im}(\chi)\sum_j\Bigg\{{\tt Im}\sum_{mn}\sum_{\alpha\beta}\frac{(G_\circ^*+w_m^*)Z^*Z}{(Z^*-w_m^*)(Z-w_n)}\frac{\langle m| i\alpha\rangle\langle m| j\beta\rangle}{\langle m|\bar m\rangle}\frac{\langle i\alpha|n\rangle\langle j\beta|n\rangle}{\langle \bar n|n\rangle}\Theta(\omega,T_j)\Bigg\}
\end{equation}
With $\Theta(\omega,T_j)=\hbar\omega\Big[\frac{1}{2}+n(\omega,T_j)\Big]$. The first summation takes over all particles in the system, i.e., $j=1,2,\cdots,N$ and accounts the total power dissipated in the $i$-th nanoparticles. The summands in which $j\neq i$, are related to the the radiative heating of {\it i}-th nanoparticle due to the radiation of the {\it j}-th particle with temperature $T_j$. In a same manner, the term $j=i$ is related to the power it lost by radiation (i.e., radiative cooling). In case where $j\neq i$ the summand reads
\begin{equation}
\label{a6}
{\tt Im}\big[\overline{{\bf E}_i^*(\omega)\cdot{{\bf P}}_i(\omega)}\big]\Big|_{j\neq i}=\frac{4\pi|Z|^2}{\omega}\Theta(\omega,T_j){\tt Im}(\chi){\tt Im}\sum_{mn}\sum_{\alpha\beta}\frac{G_\circ^*+w_m^*}{(Z^*-w_m^*)(Z-w_n)}\frac{\langle m| i\alpha\rangle\langle m| j\beta\rangle}{\langle m|\bar m\rangle}\frac{\langle i\alpha|n\rangle\langle j\beta|n\rangle}{\langle \bar n|n\rangle}.
\end{equation}
where $|Z|^2=ZZ^*$. Using $G_\circ^*+w_m^*=|Z|^2\chi-(Z^*-w_m^*)$, it is straightforward to show that the above equation equals:

\begin{eqnarray}
\label{a7}
\nonumber{\tt Im}\big[\overline{{\bf E}_i^*(\omega)\cdot{{\bf P}}_i(\omega)}\big]\Big|_{j\neq i}=\frac{4\pi|Z|^2}{\omega}\Theta(\omega,T_j){\tt Im}(\chi){\tt Im}\Bigg(&&\chi\sum_{\alpha\beta}\sum_{m}\frac{Z^*}{(Z^*-w_m^*)}\frac{\langle i\alpha|\bar m\rangle\langle m| j\beta\rangle}{\langle m|\bar m\rangle}\sum_{n}\frac{Z}{(Z-w_n)}\frac{\langle i\alpha|n\rangle\langle\bar n| j\beta\rangle}{\langle \bar n|n\rangle}\\
&&~-\sum_{\alpha\beta}\sum_{n}\frac{1}{(Z-w_n)}\frac{\langle i\alpha|n\rangle\langle\bar n| j\beta\rangle}{\langle \bar n|n\rangle}\sum_{m}\frac{\langle i\alpha|\bar m\rangle\langle m| j\beta\rangle}{\langle m|\bar m\rangle}\Bigg)
\end{eqnarray}
Here, we have used $\langle j\beta|n\rangle=\langle\bar n| j\beta\rangle$ and $\langle m| i\alpha\rangle=\langle i\alpha|\bar m\rangle$. The last summand in the second term of Eq.~(\ref{a7}) reduces to
\begin{equation}
\label{a8}
\sum_{m}\frac{\langle i\alpha|\bar m\rangle\langle m| j\beta\rangle}{\langle m|\bar m\rangle}=
\langle i\alpha|\Bigg(\sum_{m}\frac{|\bar m\rangle\langle m|}{\langle m|\bar m\rangle}\Bigg)|j\beta\rangle=\delta_{ij}\delta_{\alpha\beta}
\end{equation}
However, by assumption $j\neq i$, from which it follows that $\langle i\alpha|j\beta\rangle=0$. As a result, Eq.~(\ref{a7}) will reduces to

\begin{equation}
\label{a9}
{\tt Im}\big[\overline{{\bf E}_i^*(\omega)\cdot{{\bf P}}_i(\omega)}\big]\Big|_{j\neq i}=\frac{4\pi|Z|^2}{\omega}\Theta(\omega,T_j)\big[{\tt Im}(\chi)\big]^2\sum_{\alpha\beta}|f_{ij}(\alpha,\beta)|^2,
\end{equation}
where 
\begin{equation}
\label{a10}
f_{ij}(\alpha,\beta)=\sum_{l=1}^{3N}\frac{Z}{(Z-w_l)}\frac{\langle i\alpha|l\rangle\langle\bar l| j\beta\rangle}{\langle \bar l|l\rangle}.
\end{equation}
The summation is  taken over all elements of eigenvalues spectrum. Substituting Eq.~(\ref{a9}) into Eq.~(\ref{a1}), the radiative heating of the $i$-th particle by the $j$-th one would be
\begin{equation}
\label{a11}
\mathcal{F}_{ij}=\int_0^\infty\frac{d\omega}{2\pi}{\mathcal T}_{ij}(\omega)\Theta(\omega,T_j)
\end{equation}
with transmission coefficient
\begin{equation}
\label{a12}
{\mathcal T}_{ij}(\omega)=4|Z|^2\big[{\tt Im}(\chi)\big]^2\sum_{\alpha\beta}|f_{ij}(\alpha,\beta)|^2.
\end{equation}
We now draw our attention to the radiative cooling of the nanoparticle. Starting from Eq.~(\ref{a6}), and setting $j=i$, the only difference compared with Eq.~(\ref{a7}) is the last term will not vanish any more and we get 
\begin{equation}
\label{a13}
{\tt Im}\big[\overline{{\bf E}_i^*(\omega)\cdot{{\bf P}}_i(\omega)}\big]\Big|_{j= i}=\frac{4\pi|Z|^2}{\omega}\Theta(\omega,T_j){\tt Im}(\chi)\Bigg[{\tt Im}(\chi)\sum_{\alpha\beta}|f_{ii}(\alpha,\beta)|^2
-{\tt Im}\sum_{\alpha}\frac{f_{ii}(\alpha,\alpha)}{Z}\Bigg],
\end{equation}
Subsititing Eq.~(\ref{a13}) into Eq.~(\ref{a1}), the radiative cooling of the $i$-th particle would be
\begin{equation}
\label{a14}
\mathcal{F}_{i}=\int_0^\infty\frac{d\omega}{2\pi}{\mathcal T}_{ii}(\omega)\Theta(\omega,T_i)
\end{equation}
with cooling coeffiesent
\begin{equation}
\label{a15}
{\mathcal T}_{ii}(\omega)=4|Z|^2{\tt Im}(\chi)\Bigg[{\tt Im}(\chi)\sum_{\alpha\beta}|f_{ii}(\alpha,\beta)|^2
-{\tt Im}\sum_{\alpha}\frac{f_{ii}(\alpha,\alpha)}{Z}\Bigg],
\end{equation}
Eq.~(\ref{a12}) and (\ref{a15}) allow us to interpret the heat transfer and radiativr cooling as a summation over dipolar excitation.
\end{widetext}

\section*{reference}

\begin{thebibliography}{41}%
\makeatletter
\providecommand \@ifxundefined [1]{%
 \@ifx{#1\undefined}
}%
\providecommand \@ifnum [1]{%
 \ifnum #1\expandafter \@firstoftwo
 \else \expandafter \@secondoftwo
 \fi
}%
\providecommand \@ifx [1]{%
 \ifx #1\expandafter \@firstoftwo
 \else \expandafter \@secondoftwo
 \fi
}%
\providecommand \natexlab [1]{#1}%
\providecommand \enquote  [1]{``#1''}%
\providecommand \bibnamefont  [1]{#1}%
\providecommand \bibfnamefont [1]{#1}%
\providecommand \citenamefont [1]{#1}%
\providecommand \href@noop [0]{\@secondoftwo}%
\providecommand \href [0]{\begingroup \@sanitize@url \@href}%
\providecommand \@href[1]{\@@startlink{#1}\@@href}%
\providecommand \@@href[1]{\endgroup#1\@@endlink}%
\providecommand \@sanitize@url [0]{\catcode `\\12\catcode `\$12\catcode
  `\&12\catcode `\#12\catcode `\^12\catcode `\_12\catcode `\%12\relax}%
\providecommand \@@startlink[1]{}%
\providecommand \@@endlink[0]{}%
\providecommand \url  [0]{\begingroup\@sanitize@url \@url }%
\providecommand \@url [1]{\endgroup\@href {#1}{\urlprefix }}%
\providecommand \urlprefix  [0]{URL }%
\providecommand \Eprint [0]{\href }%
\providecommand \doibase [0]{http://dx.doi.org/}%
\providecommand \selectlanguage [0]{\@gobble}%
\providecommand \bibinfo  [0]{\@secondoftwo}%
\providecommand \bibfield  [0]{\@secondoftwo}%
\providecommand \translation [1]{[#1]}%
\providecommand \BibitemOpen [0]{}%
\providecommand \bibitemStop [0]{}%
\providecommand \bibitemNoStop [0]{.\EOS\space}%
\providecommand \EOS [0]{\spacefactor3000\relax}%
\providecommand \BibitemShut  [1]{\csname bibitem#1\endcsname}%
\let\auto@bib@innerbib\@empty
\bibitem [{\citenamefont {Song}\ \emph {et~al.}(2015)\citenamefont {Song},
  \citenamefont {Fiorino}, \citenamefont {Meyhofer},\ and\ \citenamefont
  {Reddy}}]{theorytoexperiment}%
  \BibitemOpen
  \bibfield  {author} {\bibinfo {author} {\bibfnamefont {B.}~\bibnamefont
  {Song}}, \bibinfo {author} {\bibfnamefont {A.}~\bibnamefont {Fiorino}},
  \bibinfo {author} {\bibfnamefont {E.}~\bibnamefont {Meyhofer}}, \ and\
  \bibinfo {author} {\bibfnamefont {P.}~\bibnamefont {Reddy}},\ }\href
  {\doibase 10.1063/1.4919048} {\bibfield  {journal} {\bibinfo  {journal} {AIP
  Advances}\ }\textbf {\bibinfo {volume} {5}},\ \bibinfo {pages} {053503}
  (\bibinfo {year} {2015})},\ \Eprint
  {http://arxiv.org/abs/http://dx.doi.org/10.1063/1.4919048}
  {http://dx.doi.org/10.1063/1.4919048} \BibitemShut {NoStop}%
\bibitem [{\citenamefont {Polder}\ and\ \citenamefont
  {Van~Hove}(1971)}]{VanHove}%
  \BibitemOpen
  \bibfield  {author} {\bibinfo {author} {\bibfnamefont {D.}~\bibnamefont
  {Polder}}\ and\ \bibinfo {author} {\bibfnamefont {M.}~\bibnamefont
  {Van~Hove}},\ }\href {\doibase 10.1103/PhysRevB.4.3303} {\bibfield  {journal}
  {\bibinfo  {journal} {Phys. Rev. B}\ }\textbf {\bibinfo {volume} {4}},\
  \bibinfo {pages} {3303} (\bibinfo {year} {1971})}\BibitemShut {NoStop}%
\bibitem [{\citenamefont {Narayanaswamy}\ and\ \citenamefont
  {Chen}(2008)}]{twobodysize}%
  \BibitemOpen
  \bibfield  {author} {\bibinfo {author} {\bibfnamefont {A.}~\bibnamefont
  {Narayanaswamy}}\ and\ \bibinfo {author} {\bibfnamefont {G.}~\bibnamefont
  {Chen}},\ }\href {\doibase 10.1103/PhysRevB.77.075125} {\bibfield  {journal}
  {\bibinfo  {journal} {Phys. Rev. B}\ }\textbf {\bibinfo {volume} {77}},\
  \bibinfo {pages} {075125} (\bibinfo {year} {2008})}\BibitemShut {NoStop}%
\bibitem [{\citenamefont {Sasihithlu}\ and\ \citenamefont
  {Narayanaswamy}(2014)}]{twobodysize2}%
  \BibitemOpen
  \bibfield  {author} {\bibinfo {author} {\bibfnamefont {K.}~\bibnamefont
  {Sasihithlu}}\ and\ \bibinfo {author} {\bibfnamefont {A.}~\bibnamefont
  {Narayanaswamy}},\ }\href {\doibase 10.1364/OE.22.014473} {\bibfield
  {journal} {\bibinfo  {journal} {Opt. Express}\ }\textbf {\bibinfo {volume}
  {22}},\ \bibinfo {pages} {14473} (\bibinfo {year} {2014})}\BibitemShut
  {NoStop}%
\bibitem [{\citenamefont {Wang}\ and\ \citenamefont
  {Wu}(2016)}]{sizethreebody}%
  \BibitemOpen
  \bibfield  {author} {\bibinfo {author} {\bibfnamefont {Y.}~\bibnamefont
  {Wang}}\ and\ \bibinfo {author} {\bibfnamefont {J.}~\bibnamefont {Wu}},\
  }\href {\doibase 10.1063/1.4941751} {\bibfield  {journal} {\bibinfo
  {journal} {AIP Advances}\ }\textbf {\bibinfo {volume} {6}},\ \bibinfo {pages}
  {025104} (\bibinfo {year} {2016})},\ \Eprint
  {http://arxiv.org/abs/http://dx.doi.org/10.1063/1.4941751}
  {http://dx.doi.org/10.1063/1.4941751} \BibitemShut {NoStop}%
\bibitem [{\citenamefont {Biehs}\ and\ \citenamefont
  {Agarwal}(2014)}]{spheroidslab}%
  \BibitemOpen
  \bibfield  {author} {\bibinfo {author} {\bibfnamefont {S.-A.}\ \bibnamefont
  {Biehs}}\ and\ \bibinfo {author} {\bibfnamefont {G.~S.}\ \bibnamefont
  {Agarwal}},\ }\href {\doibase 10.1103/PhysRevA.90.042510} {\bibfield
  {journal} {\bibinfo  {journal} {Phys. Rev. A}\ }\textbf {\bibinfo {volume}
  {90}},\ \bibinfo {pages} {042510} (\bibinfo {year} {2014})}\BibitemShut
  {NoStop}%
\bibitem [{\citenamefont {Choubdar}\ and\ \citenamefont
  {Nikbakht}(2016)}]{threebodyshape}%
  \BibitemOpen
  \bibfield  {author} {\bibinfo {author} {\bibfnamefont {O.~R.}\ \bibnamefont
  {Choubdar}}\ and\ \bibinfo {author} {\bibfnamefont {M.}~\bibnamefont
  {Nikbakht}},\ }\href {\doibase 10.1063/1.4964698} {\bibfield  {journal}
  {\bibinfo  {journal} {Journal of Applied Physics}\ }\textbf {\bibinfo
  {volume} {120}},\ \bibinfo {pages} {144303} (\bibinfo {year} {2016})},\
  \Eprint {http://arxiv.org/abs/http://dx.doi.org/10.1063/1.4964698}
  {http://dx.doi.org/10.1063/1.4964698} \BibitemShut {NoStop}%
\bibitem [{\citenamefont {Chapuis}\ \emph {et~al.}(2008)\citenamefont
  {Chapuis}, \citenamefont {Laroche}, \citenamefont {Volz},\ and\ \citenamefont
  {Greffet}}]{twobodymetal}%
  \BibitemOpen
  \bibfield  {author} {\bibinfo {author} {\bibfnamefont {P.-O.}\ \bibnamefont
  {Chapuis}}, \bibinfo {author} {\bibfnamefont {M.}~\bibnamefont {Laroche}},
  \bibinfo {author} {\bibfnamefont {S.}~\bibnamefont {Volz}}, \ and\ \bibinfo
  {author} {\bibfnamefont {J.-J.}\ \bibnamefont {Greffet}},\ }\href {\doibase
  10.1063/1.2931062} {\bibfield  {journal} {\bibinfo  {journal} {Applied
  Physics Letters}\ }\textbf {\bibinfo {volume} {92}},\ \bibinfo {pages}
  {201906} (\bibinfo {year} {2008})},\ \Eprint
  {http://arxiv.org/abs/http://dx.doi.org/10.1063/1.2931062}
  {http://dx.doi.org/10.1063/1.2931062} \BibitemShut {NoStop}%
\bibitem [{\citenamefont {Boriskina}\ \emph {et~al.}(2015)\citenamefont
  {Boriskina}, \citenamefont {Tong}, \citenamefont {Huang}, \citenamefont
  {Zhou}, \citenamefont {Chiloyan},\ and\ \citenamefont {Chen}}]{materialslab}%
  \BibitemOpen
  \bibfield  {author} {\bibinfo {author} {\bibfnamefont {S.}~\bibnamefont
  {Boriskina}}, \bibinfo {author} {\bibfnamefont {J.}~\bibnamefont {Tong}},
  \bibinfo {author} {\bibfnamefont {Y.}~\bibnamefont {Huang}}, \bibinfo
  {author} {\bibfnamefont {J.}~\bibnamefont {Zhou}}, \bibinfo {author}
  {\bibfnamefont {V.}~\bibnamefont {Chiloyan}}, \ and\ \bibinfo {author}
  {\bibfnamefont {G.}~\bibnamefont {Chen}},\ }\href {\doibase
  10.3390/photonics2020659} {\bibfield  {journal} {\bibinfo  {journal}
  {Photonics}\ }\textbf {\bibinfo {volume} {2}},\ \bibinfo {pages} {659â€“683}
  (\bibinfo {year} {2015})}\BibitemShut {NoStop}%
\bibitem [{\citenamefont {Ben-Abdallah}(2016)}]{manybodymagnetic}%
  \BibitemOpen
  \bibfield  {author} {\bibinfo {author} {\bibfnamefont {P.}~\bibnamefont
  {Ben-Abdallah}},\ }\href {\doibase 10.1103/PhysRevLett.116.084301} {\bibfield
   {journal} {\bibinfo  {journal} {Phys. Rev. Lett.}\ }\textbf {\bibinfo
  {volume} {116}},\ \bibinfo {pages} {084301} (\bibinfo {year}
  {2016})}\BibitemShut {NoStop}%
\bibitem [{\citenamefont {Nikbakht}(2015)}]{dynamicsNikbakht}%
  \BibitemOpen
  \bibfield  {author} {\bibinfo {author} {\bibfnamefont {M.}~\bibnamefont
  {Nikbakht}},\ }\href {http://stacks.iop.org/0295-5075/110/i=1/a=14004}
  {\bibfield  {journal} {\bibinfo  {journal} {EPL (Europhysics Letters)}\
  }\textbf {\bibinfo {volume} {110}},\ \bibinfo {pages} {14004} (\bibinfo
  {year} {2015})}\BibitemShut {NoStop}%
\bibitem [{\citenamefont {Messina}\ \emph {et~al.}(2013)\citenamefont
  {Messina}, \citenamefont {Tschikin}, \citenamefont {Biehs},\ and\
  \citenamefont {Ben-Abdallah}}]{threebodydynamicmessina}%
  \BibitemOpen
  \bibfield  {author} {\bibinfo {author} {\bibfnamefont {R.}~\bibnamefont
  {Messina}}, \bibinfo {author} {\bibfnamefont {M.}~\bibnamefont {Tschikin}},
  \bibinfo {author} {\bibfnamefont {S.-A.}\ \bibnamefont {Biehs}}, \ and\
  \bibinfo {author} {\bibfnamefont {P.}~\bibnamefont {Ben-Abdallah}},\ }\href
  {\doibase 10.1103/PhysRevB.88.104307} {\bibfield  {journal} {\bibinfo
  {journal} {Phys. Rev. B}\ }\textbf {\bibinfo {volume} {88}},\ \bibinfo
  {pages} {104307} (\bibinfo {year} {2013})}\BibitemShut {NoStop}%
\bibitem [{\citenamefont {Nagarajan}\ \emph {et~al.}(2008)\citenamefont
  {Nagarajan}, \citenamefont {Hatton}, \citenamefont {of~Colloid},
  \citenamefont {Chemistry},\ and\ \citenamefont {Meeting}}]{synthesis2}%
  \BibitemOpen
  \bibfield  {author} {\bibinfo {author} {\bibfnamefont {R.}~\bibnamefont
  {Nagarajan}}, \bibinfo {author} {\bibfnamefont {T.}~\bibnamefont {Hatton}},
  \bibinfo {author} {\bibfnamefont {A.~C. S.~D.}\ \bibnamefont {of~Colloid}},
  \bibinfo {author} {\bibfnamefont {S.}~\bibnamefont {Chemistry}}, \ and\
  \bibinfo {author} {\bibfnamefont {A.~C.~S.}\ \bibnamefont {Meeting}},\ }\href
  {https://books.google.com/books?id=W07xAAAAMAAJ} {\emph {\bibinfo {title}
  {Nanoparticles: Synthesis, Stabilization, Passivation, and
  Functionalization}}},\ ACS symposium series\ (\bibinfo  {publisher} {American
  Chemical Society},\ \bibinfo {year} {2008})\BibitemShut {NoStop}%
\bibitem [{\citenamefont {Nejo}(2007)}]{fabrication}%
  \BibitemOpen
  \bibfield  {author} {\bibinfo {author} {\bibfnamefont {H.}~\bibnamefont
  {Nejo}},\ }\href {https://books.google.com/books?id=wfFDAAAAQBAJ} {\emph
  {\bibinfo {title} {Nanostructures: Fabrication and Analysis}}},\ NanoScience
  and Technology\ (\bibinfo  {publisher} {Springer Berlin Heidelberg},\
  \bibinfo {year} {2007})\BibitemShut {NoStop}%
\bibitem [{\citenamefont {Nikbakht}(2014)}]{manybodynikbakht}%
  \BibitemOpen
  \bibfield  {author} {\bibinfo {author} {\bibfnamefont {M.}~\bibnamefont
  {Nikbakht}},\ }\href {\doibase 10.1063/1.4894622} {\bibfield  {journal}
  {\bibinfo  {journal} {Journal of Applied Physics}\ }\textbf {\bibinfo
  {volume} {116}},\ \bibinfo {pages} {094307} (\bibinfo {year} {2014})},\
  \Eprint {http://arxiv.org/abs/http://dx.doi.org/10.1063/1.4894622}
  {http://dx.doi.org/10.1063/1.4894622} \BibitemShut {NoStop}%
\bibitem [{\citenamefont {Ben-Abdallah}\ \emph {et~al.}(2011)\citenamefont
  {Ben-Abdallah}, \citenamefont {Biehs},\ and\ \citenamefont
  {Joulain}}]{many-bodyben}%
  \BibitemOpen
  \bibfield  {author} {\bibinfo {author} {\bibfnamefont {P.}~\bibnamefont
  {Ben-Abdallah}}, \bibinfo {author} {\bibfnamefont {S.-A.}\ \bibnamefont
  {Biehs}}, \ and\ \bibinfo {author} {\bibfnamefont {K.}~\bibnamefont
  {Joulain}},\ }\href {\doibase 10.1103/PhysRevLett.107.114301} {\bibfield
  {journal} {\bibinfo  {journal} {Phys. Rev. Lett.}\ }\textbf {\bibinfo
  {volume} {107}},\ \bibinfo {pages} {114301} (\bibinfo {year}
  {2011})}\BibitemShut {NoStop}%
\bibitem [{\citenamefont {Dong}\ \emph
  {et~al.}(2017{\natexlab{a}})\citenamefont {Dong}, \citenamefont {Zhao},\ and\
  \citenamefont {Liu}}]{manybodyzhao}%
  \BibitemOpen
  \bibfield  {author} {\bibinfo {author} {\bibfnamefont {J.}~\bibnamefont
  {Dong}}, \bibinfo {author} {\bibfnamefont {J.}~\bibnamefont {Zhao}}, \ and\
  \bibinfo {author} {\bibfnamefont {L.}~\bibnamefont {Liu}},\ }\href {\doibase
  10.1103/PhysRevB.95.125411} {\bibfield  {journal} {\bibinfo  {journal} {Phys.
  Rev. B}\ }\textbf {\bibinfo {volume} {95}},\ \bibinfo {pages} {125411}
  (\bibinfo {year} {2017}{\natexlab{a}})}\BibitemShut {NoStop}%
\bibitem [{\citenamefont {Latella}\ \emph
  {et~al.}(2017{\natexlab{a}})\citenamefont {Latella}, \citenamefont
  {Ben-Abdallah}, \citenamefont {Biehs}, \citenamefont {Antezza},\ and\
  \citenamefont {Messina}}]{manybodyslab}%
  \BibitemOpen
  \bibfield  {author} {\bibinfo {author} {\bibfnamefont {I.}~\bibnamefont
  {Latella}}, \bibinfo {author} {\bibfnamefont {P.}~\bibnamefont
  {Ben-Abdallah}}, \bibinfo {author} {\bibfnamefont {S.-A.}\ \bibnamefont
  {Biehs}}, \bibinfo {author} {\bibfnamefont {M.}~\bibnamefont {Antezza}}, \
  and\ \bibinfo {author} {\bibfnamefont {R.}~\bibnamefont {Messina}},\ }\href
  {\doibase 10.1103/PhysRevB.95.205404} {\bibfield  {journal} {\bibinfo
  {journal} {Phys. Rev. B}\ }\textbf {\bibinfo {volume} {95}},\ \bibinfo
  {pages} {205404} (\bibinfo {year} {2017}{\natexlab{a}})}\BibitemShut
  {NoStop}%
\bibitem [{\citenamefont {Ben-Abdallah}\ \emph {et~al.}(2013)\citenamefont
  {Ben-Abdallah}, \citenamefont {Messina}, \citenamefont {Biehs}, \citenamefont
  {Tschikin}, \citenamefont {Joulain},\ and\ \citenamefont
  {Henkel}}]{manybodyheatdiffusion}%
  \BibitemOpen
  \bibfield  {author} {\bibinfo {author} {\bibfnamefont {P.}~\bibnamefont
  {Ben-Abdallah}}, \bibinfo {author} {\bibfnamefont {R.}~\bibnamefont
  {Messina}}, \bibinfo {author} {\bibfnamefont {S.-A.}\ \bibnamefont {Biehs}},
  \bibinfo {author} {\bibfnamefont {M.}~\bibnamefont {Tschikin}}, \bibinfo
  {author} {\bibfnamefont {K.}~\bibnamefont {Joulain}}, \ and\ \bibinfo
  {author} {\bibfnamefont {C.}~\bibnamefont {Henkel}},\ }\href {\doibase
  10.1103/PhysRevLett.111.174301} {\bibfield  {journal} {\bibinfo  {journal}
  {Phys. Rev. Lett.}\ }\textbf {\bibinfo {volume} {111}},\ \bibinfo {pages}
  {174301} (\bibinfo {year} {2013})}\BibitemShut {NoStop}%
\bibitem [{\citenamefont {Cherny}\ \emph {et~al.}(2011)\citenamefont {Cherny},
  \citenamefont {Anitas}, \citenamefont {Osipov},\ and\ \citenamefont
  {Kuklin}}]{fractalscattering}%
  \BibitemOpen
  \bibfield  {author} {\bibinfo {author} {\bibfnamefont {A.~Y.}\ \bibnamefont
  {Cherny}}, \bibinfo {author} {\bibfnamefont {E.~M.}\ \bibnamefont {Anitas}},
  \bibinfo {author} {\bibfnamefont {V.~A.}\ \bibnamefont {Osipov}}, \ and\
  \bibinfo {author} {\bibfnamefont {A.~I.}\ \bibnamefont {Kuklin}},\ }\href
  {\doibase 10.1103/PhysRevE.84.036203} {\bibfield  {journal} {\bibinfo
  {journal} {Phys. Rev. E}\ }\textbf {\bibinfo {volume} {84}},\ \bibinfo
  {pages} {036203} (\bibinfo {year} {2011})}\BibitemShut {NoStop}%
\bibitem [{\citenamefont {Stockman}(1997)}]{localizemodes}%
  \BibitemOpen
  \bibfield  {author} {\bibinfo {author} {\bibfnamefont {M.~I.}\ \bibnamefont
  {Stockman}},\ }\href {\doibase 10.1103/PhysRevE.56.6494} {\bibfield
  {journal} {\bibinfo  {journal} {Phys. Rev. E}\ }\textbf {\bibinfo {volume}
  {56}},\ \bibinfo {pages} {6494} (\bibinfo {year} {1997})}\BibitemShut
  {NoStop}%
\bibitem [{\citenamefont {Ben-Abdallah}\ \emph {et~al.}(2008)\citenamefont
  {Ben-Abdallah}, \citenamefont {Joulain}, \citenamefont {Drevillon},\ and\
  \citenamefont {Le~Goff}}]{arrayplasmonic}%
  \BibitemOpen
  \bibfield  {author} {\bibinfo {author} {\bibfnamefont {P.}~\bibnamefont
  {Ben-Abdallah}}, \bibinfo {author} {\bibfnamefont {K.}~\bibnamefont
  {Joulain}}, \bibinfo {author} {\bibfnamefont {J.}~\bibnamefont {Drevillon}},
  \ and\ \bibinfo {author} {\bibfnamefont {C.}~\bibnamefont {Le~Goff}},\ }\href
  {\doibase 10.1103/PhysRevB.77.075417} {\bibfield  {journal} {\bibinfo
  {journal} {Phys. Rev. B}\ }\textbf {\bibinfo {volume} {77}},\ \bibinfo
  {pages} {075417} (\bibinfo {year} {2008})}\BibitemShut {NoStop}%
\bibitem [{\citenamefont {Ben-Abdallah}(2006)}]{arraynanofluid}%
  \BibitemOpen
  \bibfield  {author} {\bibinfo {author} {\bibfnamefont {P.}~\bibnamefont
  {Ben-Abdallah}},\ }\href {\doibase 10.1063/1.2349857} {\bibfield  {journal}
  {\bibinfo  {journal} {Applied Physics Letters}\ }\textbf {\bibinfo {volume}
  {89}},\ \bibinfo {pages} {113117} (\bibinfo {year} {2006})},\ \Eprint
  {http://arxiv.org/abs/http://dx.doi.org/10.1063/1.2349857}
  {http://dx.doi.org/10.1063/1.2349857} \BibitemShut {NoStop}%
\bibitem [{\citenamefont {Latella}\ \emph
  {et~al.}(2017{\natexlab{b}})\citenamefont {Latella}, \citenamefont
  {Ben-Abdallah}, \citenamefont {Biehs}, \citenamefont {Antezza},\ and\
  \citenamefont {Messina}}]{momentumheattransfermanybodyplanar}%
  \BibitemOpen
  \bibfield  {author} {\bibinfo {author} {\bibfnamefont {I.}~\bibnamefont
  {Latella}}, \bibinfo {author} {\bibfnamefont {P.}~\bibnamefont
  {Ben-Abdallah}}, \bibinfo {author} {\bibfnamefont {S.-A.}\ \bibnamefont
  {Biehs}}, \bibinfo {author} {\bibfnamefont {M.}~\bibnamefont {Antezza}}, \
  and\ \bibinfo {author} {\bibfnamefont {R.}~\bibnamefont {Messina}},\ }\href
  {\doibase 10.1103/PhysRevB.95.205404} {\bibfield  {journal} {\bibinfo
  {journal} {Phys. Rev. B}\ }\textbf {\bibinfo {volume} {95}},\ \bibinfo
  {pages} {205404} (\bibinfo {year} {2017}{\natexlab{b}})}\BibitemShut
  {NoStop}%
\bibitem [{\citenamefont {Latella}\ \emph
  {et~al.}(2017{\natexlab{c}})\citenamefont {Latella}, \citenamefont {Biehs},
  \citenamefont {Messina}, \citenamefont {Rodriguez},\ and\ \citenamefont
  {Ben-Abdallah}}]{ballesticslab}%
  \BibitemOpen
  \bibfield  {author} {\bibinfo {author} {\bibfnamefont {I.}~\bibnamefont
  {Latella}}, \bibinfo {author} {\bibfnamefont {S.-A.}\ \bibnamefont {Biehs}},
  \bibinfo {author} {\bibfnamefont {R.}~\bibnamefont {Messina}}, \bibinfo
  {author} {\bibfnamefont {A.~W.}\ \bibnamefont {Rodriguez}}, \ and\ \bibinfo
  {author} {\bibfnamefont {P.}~\bibnamefont {Ben-Abdallah}},\ }\href@noop {}
  {\enquote {\bibinfo {title} {Ballistic near-field heat transport in dense
  many-body systems},}\ } (\bibinfo {year} {2017}{\natexlab{c}}),\ \Eprint
  {http://arxiv.org/abs/arXiv:1705.07148} {arXiv:1705.07148} \BibitemShut
  {NoStop}%
\bibitem [{\citenamefont {Dong}\ \emph
  {et~al.}(2017{\natexlab{b}})\citenamefont {Dong}, \citenamefont {Zhao},\ and\
  \citenamefont {Liu}}]{htbetweencluster}%
  \BibitemOpen
  \bibfield  {author} {\bibinfo {author} {\bibfnamefont {J.}~\bibnamefont
  {Dong}}, \bibinfo {author} {\bibfnamefont {J.}~\bibnamefont {Zhao}}, \ and\
  \bibinfo {author} {\bibfnamefont {L.}~\bibnamefont {Liu}},\ }\href {\doibase
  http://dx.doi.org/10.1016/j.jqsrt.2016.10.015} {\bibfield  {journal}
  {\bibinfo  {journal} {Journal of Quantitative Spectroscopy and Radiative
  Transfer}\ }\textbf {\bibinfo {volume} {197}},\ \bibinfo {pages} {114 }
  (\bibinfo {year} {2017}{\natexlab{b}})},\ \bibinfo {note} {the Eight
  International Symposium on Radiative Transfer}\BibitemShut {NoStop}%
\bibitem [{\citenamefont {Latella}\ and\ \citenamefont
  {Ben-Abdallah}(2017)}]{magnetoplasmonicchain}%
  \BibitemOpen
  \bibfield  {author} {\bibinfo {author} {\bibfnamefont {I.}~\bibnamefont
  {Latella}}\ and\ \bibinfo {author} {\bibfnamefont {P.}~\bibnamefont
  {Ben-Abdallah}},\ }\href {\doibase 10.1103/PhysRevLett.118.173902} {\bibfield
   {journal} {\bibinfo  {journal} {Phys. Rev. Lett.}\ }\textbf {\bibinfo
  {volume} {118}},\ \bibinfo {pages} {173902} (\bibinfo {year}
  {2017})}\BibitemShut {NoStop}%
\bibitem [{\citenamefont {Vicsek}(1992)}]{vicsek1992fractal}%
  \BibitemOpen
  \bibfield  {author} {\bibinfo {author} {\bibfnamefont {T.}~\bibnamefont
  {Vicsek}},\ }\href {https://books.google.com/books?id=InnD-GTUi0gC} {\emph
  {\bibinfo {title} {Fractal Growth Phenomena}}}\ (\bibinfo  {publisher} {World
  Scientific},\ \bibinfo {year} {1992})\BibitemShut {NoStop}%
\bibitem [{\citenamefont {Shalaev}(1999)}]{opticalproperties1}%
  \BibitemOpen
  \bibfield  {author} {\bibinfo {author} {\bibfnamefont {V.}~\bibnamefont
  {Shalaev}},\ }\href {https://books.google.com/books?id=tBzEMlmBElMC} {\emph
  {\bibinfo {title} {Nonlinear Optics of Random Media: Fractal Composites and
  Metal-Dielectric Films}}},\ Springer Tracts in Modern Physics\ (\bibinfo
  {publisher} {Springer Berlin Heidelberg},\ \bibinfo {year}
  {1999})\BibitemShut {NoStop}%
\bibitem [{\citenamefont {Mandelbrot}(1982)}]{mandelbrot}%
  \BibitemOpen
  \bibfield  {author} {\bibinfo {author} {\bibfnamefont {B.}~\bibnamefont
  {Mandelbrot}},\ }\href {https://books.google.com/books?id=0R2LkE3N7-oC}
  {\emph {\bibinfo {title} {The Fractal Geometry of Nature}}}\ (\bibinfo
  {publisher} {Henry Holt and Company},\ \bibinfo {year} {1982})\BibitemShut
  {NoStop}%
\bibitem [{\citenamefont {Witten}\ and\ \citenamefont
  {Sander}(1983)}]{wittensander}%
  \BibitemOpen
  \bibfield  {author} {\bibinfo {author} {\bibfnamefont {T.~A.}\ \bibnamefont
  {Witten}}\ and\ \bibinfo {author} {\bibfnamefont {L.~M.}\ \bibnamefont
  {Sander}},\ }\href {\doibase 10.1103/PhysRevB.27.5686} {\bibfield  {journal}
  {\bibinfo  {journal} {Phys. Rev. B}\ }\textbf {\bibinfo {volume} {27}},\
  \bibinfo {pages} {5686} (\bibinfo {year} {1983})}\BibitemShut {NoStop}%
\bibitem [{\citenamefont {Meakin}(1983)}]{meakinfractal2}%
  \BibitemOpen
  \bibfield  {author} {\bibinfo {author} {\bibfnamefont {P.}~\bibnamefont
  {Meakin}},\ }\href {\doibase 10.1103/PhysRevLett.51.1119} {\bibfield
  {journal} {\bibinfo  {journal} {Phys. Rev. Lett.}\ }\textbf {\bibinfo
  {volume} {51}},\ \bibinfo {pages} {1119} (\bibinfo {year}
  {1983})}\BibitemShut {NoStop}%
\bibitem [{\citenamefont {Shalaev}(2003)}]{opticalproperties}%
  \BibitemOpen
  \bibfield  {author} {\bibinfo {author} {\bibfnamefont {V.}~\bibnamefont
  {Shalaev}},\ }\href {https://books.google.com/books?id=lXMKBwAAQBAJ} {\emph
  {\bibinfo {title} {Optical Properties of Nanostructured Random Media}}},\
  Topics in Applied Physics\ (\bibinfo  {publisher} {Springer Berlin
  Heidelberg},\ \bibinfo {year} {2003})\BibitemShut {NoStop}%
\bibitem [{\citenamefont {Nikbakht}\ and\ \citenamefont
  {Mahdieh}(2011)}]{nikbakht3}%
  \BibitemOpen
  \bibfield  {author} {\bibinfo {author} {\bibfnamefont {M.}~\bibnamefont
  {Nikbakht}}\ and\ \bibinfo {author} {\bibfnamefont {M.~H.}\ \bibnamefont
  {Mahdieh}},\ }\href {\doibase 10.1021/jp1088542} {\bibfield  {journal}
  {\bibinfo  {journal} {The Journal of Physical Chemistry C}\ }\textbf
  {\bibinfo {volume} {115}},\ \bibinfo {pages} {1561} (\bibinfo {year}
  {2011})},\ \Eprint {http://arxiv.org/abs/http://dx.doi.org/10.1021/jp1088542}
  {http://dx.doi.org/10.1021/jp1088542} \BibitemShut {NoStop}%
\bibitem [{\citenamefont {Stockman}\ \emph {et~al.}(1996)\citenamefont
  {Stockman}, \citenamefont {Pandey},\ and\ \citenamefont
  {George}}]{localizationstockman}%
  \BibitemOpen
  \bibfield  {author} {\bibinfo {author} {\bibfnamefont {M.~I.}\ \bibnamefont
  {Stockman}}, \bibinfo {author} {\bibfnamefont {L.~N.}\ \bibnamefont
  {Pandey}}, \ and\ \bibinfo {author} {\bibfnamefont {T.~F.}\ \bibnamefont
  {George}},\ }\href {\doibase 10.1103/PhysRevB.53.2183} {\bibfield  {journal}
  {\bibinfo  {journal} {Phys. Rev. B}\ }\textbf {\bibinfo {volume} {53}},\
  \bibinfo {pages} {2183} (\bibinfo {year} {1996})}\BibitemShut {NoStop}%
\bibitem [{\citenamefont {Sherry}\ \emph {et~al.}(2005)\citenamefont {Sherry},
  \citenamefont {Chang}, \citenamefont {Schatz}, \citenamefont {Van~Duyne},
  \citenamefont {Wiley},\ and\ \citenamefont {Xia}}]{localizedsurfaceplasmon}%
  \BibitemOpen
  \bibfield  {author} {\bibinfo {author} {\bibfnamefont {L.~J.}\ \bibnamefont
  {Sherry}}, \bibinfo {author} {\bibfnamefont {S.-H.}\ \bibnamefont {Chang}},
  \bibinfo {author} {\bibfnamefont {G.~C.}\ \bibnamefont {Schatz}}, \bibinfo
  {author} {\bibfnamefont {R.~P.}\ \bibnamefont {Van~Duyne}}, \bibinfo {author}
  {\bibfnamefont {B.~J.}\ \bibnamefont {Wiley}}, \ and\ \bibinfo {author}
  {\bibfnamefont {Y.}~\bibnamefont {Xia}},\ }\href {\doibase 10.1021/nl0515753}
  {\bibfield  {journal} {\bibinfo  {journal} {Nano Letters}\ }\textbf {\bibinfo
  {volume} {5}},\ \bibinfo {pages} {2034} (\bibinfo {year} {2005})},\ \bibinfo
  {note} {pMID: 16218733},\ \Eprint
  {http://arxiv.org/abs/http://dx.doi.org/10.1021/nl0515753}
  {http://dx.doi.org/10.1021/nl0515753} \BibitemShut {NoStop}%
\bibitem [{\citenamefont {Blumen}\ \emph {et~al.}(2004)\citenamefont {Blumen},
  \citenamefont {von Ferber}, \citenamefont {Jurjiu},\ and\ \citenamefont
  {Koslowski}}]{vicsekfractal}%
  \BibitemOpen
  \bibfield  {author} {\bibinfo {author} {\bibfnamefont {A.}~\bibnamefont
  {Blumen}}, \bibinfo {author} {\bibfnamefont {C.}~\bibnamefont {von Ferber}},
  \bibinfo {author} {\bibfnamefont {A.}~\bibnamefont {Jurjiu}}, \ and\ \bibinfo
  {author} {\bibfnamefont {T.}~\bibnamefont {Koslowski}},\ }\href {\doibase
  10.1021/ma034553g} {\bibfield  {journal} {\bibinfo  {journal}
  {Macromolecules}\ }\textbf {\bibinfo {volume} {37}},\ \bibinfo {pages} {638}
  (\bibinfo {year} {2004})},\ \Eprint
  {http://arxiv.org/abs/http://dx.doi.org/10.1021/ma034553g}
  {http://dx.doi.org/10.1021/ma034553g} \BibitemShut {NoStop}%
\bibitem [{\citenamefont {Albaladejo}\ \emph {et~al.}(2010)\citenamefont
  {Albaladejo}, \citenamefont {G\'{o}mez-Medina}, \citenamefont
  {Froufe-P\'{e}rez}, \citenamefont {Marinchio}, \citenamefont {Carminati},
  \citenamefont {Torrado}, \citenamefont {Armelles}, \citenamefont
  {Garc\'{i}a-Mart\'{i}n},\ and\ \citenamefont
  {S\'{a}enz}}]{radiationcorrection}%
  \BibitemOpen
  \bibfield  {author} {\bibinfo {author} {\bibfnamefont {S.}~\bibnamefont
  {Albaladejo}}, \bibinfo {author} {\bibfnamefont {R.}~\bibnamefont
  {G\'{o}mez-Medina}}, \bibinfo {author} {\bibfnamefont {L.~S.}\ \bibnamefont
  {Froufe-P\'{e}rez}}, \bibinfo {author} {\bibfnamefont {H.}~\bibnamefont
  {Marinchio}}, \bibinfo {author} {\bibfnamefont {R.}~\bibnamefont
  {Carminati}}, \bibinfo {author} {\bibfnamefont {J.~F.}\ \bibnamefont
  {Torrado}}, \bibinfo {author} {\bibfnamefont {G.}~\bibnamefont {Armelles}},
  \bibinfo {author} {\bibfnamefont {A.}~\bibnamefont {Garc\'{i}a-Mart\'{i}n}},
  \ and\ \bibinfo {author} {\bibfnamefont {J.~J.}\ \bibnamefont {S\'{a}enz}},\
  }\href {\doibase 10.1364/OE.18.003556} {\bibfield  {journal} {\bibinfo
  {journal} {Opt. Express}\ }\textbf {\bibinfo {volume} {18}},\ \bibinfo
  {pages} {3556} (\bibinfo {year} {2010})}\BibitemShut {NoStop}%
\bibitem [{\citenamefont {Markel}\ \emph {et~al.}(1991)\citenamefont {Markel},
  \citenamefont {Muratov}, \citenamefont {Stockman},\ and\ \citenamefont
  {George}}]{ZZmarkel}%
  \BibitemOpen
  \bibfield  {author} {\bibinfo {author} {\bibfnamefont {V.~A.}\ \bibnamefont
  {Markel}}, \bibinfo {author} {\bibfnamefont {L.~S.}\ \bibnamefont {Muratov}},
  \bibinfo {author} {\bibfnamefont {M.~I.}\ \bibnamefont {Stockman}}, \ and\
  \bibinfo {author} {\bibfnamefont {T.~F.}\ \bibnamefont {George}},\ }\href
  {\doibase 10.1103/PhysRevB.43.8183} {\bibfield  {journal} {\bibinfo
  {journal} {Phys. Rev. B}\ }\textbf {\bibinfo {volume} {43}},\ \bibinfo
  {pages} {8183} (\bibinfo {year} {1991})}\BibitemShut {NoStop}%
\bibitem [{\citenamefont {Horn}\ and\ \citenamefont
  {Johnson}(1990)}]{hornmatrix}%
  \BibitemOpen
  \bibfield  {author} {\bibinfo {author} {\bibfnamefont {R.}~\bibnamefont
  {Horn}}\ and\ \bibinfo {author} {\bibfnamefont {C.}~\bibnamefont {Johnson}},\
  }\href {https://books.google.com/books?id=PlYQN0ypTwEC} {\emph {\bibinfo
  {title} {Matrix Analysis}}}\ (\bibinfo  {publisher} {Cambridge University
  Press},\ \bibinfo {year} {1990})\BibitemShut {NoStop}%
\bibitem [{\citenamefont {Raki\'{c}}\ \emph {et~al.}(1998)\citenamefont
  {Raki\'{c}}, \citenamefont {Djuri\v{s}i\'{c}}, \citenamefont {Elazar},\ and\
  \citenamefont {Majewski}}]{Rakicepsilon}%
  \BibitemOpen
  \bibfield  {author} {\bibinfo {author} {\bibfnamefont {A.~D.}\ \bibnamefont
  {Raki\'{c}}}, \bibinfo {author} {\bibfnamefont {A.~B.}\ \bibnamefont
  {Djuri\v{s}i\'{c}}}, \bibinfo {author} {\bibfnamefont {J.~M.}\ \bibnamefont
  {Elazar}}, \ and\ \bibinfo {author} {\bibfnamefont {M.~L.}\ \bibnamefont
  {Majewski}},\ }\href {\doibase 10.1364/AO.37.005271} {\bibfield  {journal}
  {\bibinfo  {journal} {Appl. Opt.}\ }\textbf {\bibinfo {volume} {37}},\
  \bibinfo {pages} {5271} (\bibinfo {year} {1998})}\BibitemShut {NoStop}%
\end{thebibliography}
%

\end{document}